\documentclass[aps,prx,superscriptaddress,notitlepage,twocolumn,longbibliography]{revtex4-1}

\usepackage{hyperref}
\hypersetup{
  colorlinks=true,
  linkcolor=blue,
  urlcolor=cyan,
}

\usepackage{physics} 
\usepackage{amssymb} 
\usepackage{bm} 
\usepackage{braket} 
\usepackage{mathtools} 
\usepackage{graphicx} 

\renewcommand{\t}{\text} 
\newcommand{\f}[2]{\dfrac{#1}{#2}} 
\newcommand{\p}[1]{\left(#1\right)} 
\renewcommand{\sp}[1]{\left[#1\right]} 
\renewcommand{\set}[1]{\left\{#1\right\}} 
\renewcommand{\c}{\cdot} 
\renewcommand{\d}{\text{d}} 
\renewcommand{\v}{\bm} 
\newcommand{\uv}[1]{\hat{\bm #1}} 
\newcommand{\bk}{\Braket}
\renewcommand{\ket}{\Ket}
\renewcommand{\bra}{\Bra}

\newcommand{\B}{\mathcal{B}}
\newcommand{\E}{\mathcal{E}}
\newcommand{\G}{\mathcal{G}}
\newcommand{\I}{\mathcal{I}}
\newcommand{\J}{\mathcal{J}}
\renewcommand{\L}{\mathcal{L}}
\renewcommand{\O}{\mathcal{O}}
\renewcommand{\P}{\mathcal{P}}
\newcommand{\Q}{\mathcal{Q}}
\renewcommand{\S}{\mathcal{S}}

\newcommand{\z}{\text{z}}
\newcommand{\x}{\text{x}}
\newcommand{\y}{\text{y}}

\newcommand{\up}{\uparrow}
\newcommand{\dn}{\downarrow}

\newcommand{\JILA}{JILA, National Institute of Standards and Technology and Department of Physics, University of Colorado, Boulder, CO, 80309, USA}
\newcommand{\CTQM}{Center for Theory of Quantum Matter, University of Colorado, Boulder, CO, 80309, USA}
\newcommand{\contrib}{\thanks{Authors P.H.~and M.A.P.~contributed equally to this work.}}

\begin{document}

\title{Engineering spin squeezing in a 3D optical lattice with interacting spin-orbit-coupled fermions}

\author{P.~He} \contrib
\author{M.~A.~Perlin} \contrib
\author{S.~R.~Muleady}
\author{R.~J.~Lewis-Swan}
\affiliation{\JILA}
\affiliation{\CTQM}
\author{R.~B.~Hutson}
\author{J.~Ye}
\affiliation{\JILA}
\author{A.~M.~Rey}
\affiliation{\JILA}
\affiliation{\CTQM}

\begin{abstract}
One of the most important tasks in modern quantum science is to coherently control and entangle many-body systems, and to subsequently use these systems to realize powerful quantum technologies such as quantum-enhanced sensors.
However, many-body entangled states are difficult to prepare and preserve since internal dynamics and external noise rapidly degrade any useful entanglement.
Here, we introduce a protocol that counterintuitively exploits inhomogeneities, a typical source of dephasing in a many-body system, in combination with interactions to generate metrologically useful and robust many-body entangled states.
Motivated by current limitations in state-of-the-art three-dimensional (3D) optical lattice clocks (OLCs) operating at quantum degeneracy, we use local interactions in a Hubbard model with spin-orbit coupling to achieve a spin-locking effect.
In addition to prolonging inter-particle spin coherence, spin-locking transforms the dephasing effect of spin-orbit coupling into a collective spin-squeezing process that can be further enhanced by applying a modulated drive.
Our protocol is fully compatible with state-of-the-art 3D OLC interrogation schemes and may be used to improve their sensitivity, which is currently limited by the intrinsic quantum noise of independent atoms.
We demonstrate that even with realistic experimental imperfections, our protocol may generate $\sim10$--$14$ dB of spin squeezing in $\sim1$ second with $\sim10^2$--$10^4$ atoms.
This capability allows OLCs to enter a new era of quantum enhanced sensing using correlated quantum states of driven non-equilibrium systems.
\end{abstract}


\maketitle

\section{Introduction}

A major frontier of contemporary physics is the understanding of non-equilibrium behaviors of many-body quantum systems, and the application of these behaviors toward the development of novel quantum technologies with untapped capabilities\cite{eisert2015quantum}.
To this end, ultracold atomic, molecular, and optical systems are ideal platforms for studying unexplored regimes of many-body physics due to their clean preparation and readout, high controllability, and long coherence times\cite{bloch2008manybody, gross2017quantum}.
The exquisite capabilities of these systems have pushed the frontiers of metrology, quantum simulation, and quantum information science.

Optical lattice clocks in particular have seen some of the most impressive developments in recent years, reaching record levels of precision ($\sim 3\times 10^{-19}$)\cite{campbell2017fermidegenerate, marti2018imaging} and accuracy ($\sim 1\times 10^{-18}$)\cite{bloom2014optical, mcgrew2018atomic}.
These advancements required important breakthroughs, including the capability to cool and trap fermionic alkaline-earth atoms in spin-insensitive potentials\cite{takamoto2003spectroscopy, barber2006direct, ye2008quantum}; the development of ultracoherent lasers\cite{kessler2012sub40mhzlinewidth, cole2013tenfold, matei2017mu} to fully exploit an ultranarrow clock transition\cite{ludlow2015optical}; the detailed characterization of inter-atomic interactions\cite{scazza2014observation, cappellini2014direct, zhang2014spectroscopic}; and, more recently, the preparation of a quantum degenerate gas in a three-dimensional (3D) optical lattice\cite{campbell2017fermidegenerate, marti2018imaging, goban2018emergence}.
Nonetheless, all improvements in sensing capabilities to date have been based on single-particle control of internal atomic degrees of freedom.
Such strategies will eventually have diminishing returns due to practical difficulties in (i) suppressing decoherence from external (motional) degrees of freedom, and (ii) interrogating more particles without additional systematic errors from interactions\cite{martin2013quantum, ludlow2015optical, marti2018imaging}.

Pushing beyond the current independent-particle paradigm requires leveraging many-body quantum correlations.
Entangled states such as spin-squeezed states\cite{kitagawa1993squeezed, wineland1992spin, ma2011quantum} can enhance measurement sensitivity, i.e.~the uncertainty $\Delta\theta$ in the estimation of a parameter $\theta$, below the standard quantum limit $\Delta\theta\sim1/\sqrt{N}$ for $N$ uncorrelated particles\cite{itano1993quantum, degen2017quantum}.
The major challenge for progress in this direction is that generating entanglement requires interactions, which are generally undesirable because they degrade atomic coherence, thereby limiting clock performance\cite{swallows2011suppression, martin2013quantum, rey2014probing, ludlow2011coldcollisionshift, lemke2011wave, ludlow2015optical}.
In fact, the most precise and accurate optical lattice clocks were designed to operate with fermionic atoms in identical nuclear and electronic states to suppress collisional decoherence\cite{campbell2009probing, swallows2011suppression, campbell2017fermidegenerate}, as identical fermions cannot interact via the otherwise dominant ($s$-wave) collisions at ultracold temperatures.
However, an initially spin-polarized Fermi gas still exhibits interactions at later times due to spin-orbit coupling (SOC) that is induced by the laser that drives the clock transition (i.e.~the ``clock laser'')\cite{wall2016synthetic, kolkowitz2016spinorbitcoupled, livi2016synthetic, bromley2018dynamics}.
Specifically, the momentum kick imparted by this laser imprints a position-dependent phase that induces inhomogeneous spin precession and generates spin dephasing, thereby making atoms distinguishable and vulnerable to collisions.
While a deep lattice can suppress SOC, it also intensifies the light scattering which currently limits the coherence time of the clock\cite{dorscher2018latticeinduced, goban2018emergence, hutson2019engineering}.

In this work, we describe a scheme that can lead to metrological advances in state-of-the-art optical lattice clocks through direct use of quantum entanglement by harnessing the interplay between nominally undesirable collisions and SOC.
This scheme is made possible in the weak SOC regime by the formation of an interaction-energy gap that suppresses the SOC-induced population transfer from the exchange-symmetric Dicke manifold (spanned by spin-polarized, and thus non-interacting states) to the remainder of Hilbert space.
Interactions thereby prolong inter-particle spin coherence through a spin-locking effect, which additionally transforms the dephasing effect of SOC into a collective spin squeezing process.
To generate spin squeezing, our protocol only requires the capability to fix (i) the orientation of the clock laser and (ii) the optical lattice depth.
These controls are straightforward to incorporate into current 3D clock interrogation sequences without sacrificing atom numbers or coherence times.
Additionally, we show that by applying a modulated drive from the clock laser, one can further prepare states that saturate the Heisenberg limit $\Delta\theta\sim1/N$ for phase sensitivity\cite{kitagawa1993squeezed, ma2011quantum, degen2017quantum}.
This capability mirrors efforts in other settings, such as nitrogen-vacancy centers in diamond\cite{bauch2018ultralong, aiello2013compositepulse} and trapped ions\cite{burd2019quantum}, to enhance quantum metrology through the use of driven non-equilibrium phenomena.

Despite an abundance of proof-of-principle experiments with entangled states\cite{degen2017quantum, pezze2018quantum}, so far only the remarkable example of LIGO\cite{aasi2013enhanced, abbott2016gw150914} has demonstrated a quantum advantage in a state-of-the-art quantum sensing or measurement system.
The new generation of 3D optical lattice systems have fully quantized motional degrees of freedom\cite{campbell2017fermidegenerate}, allowing for precise control of collisional interactions.
We demonstrate how these interactions can naturally give rise to metrologically useful correlated many-body fermionic states, opening a path to not only generate entanglement, but also harness it to achieve a quantum advantage in a world-class sensor.
Such an advance will ultimately deliver gains to real-world applications including timekeeping, navigation, telecommunication, and our understanding of the fundamental laws of nature\cite{safronova2018search}.

\section{Spin squeezing with the Fermi-Hubbard model}
\label{sec:theory}

\begin{figure*}
\centering
\includegraphics[width=0.6\textwidth]{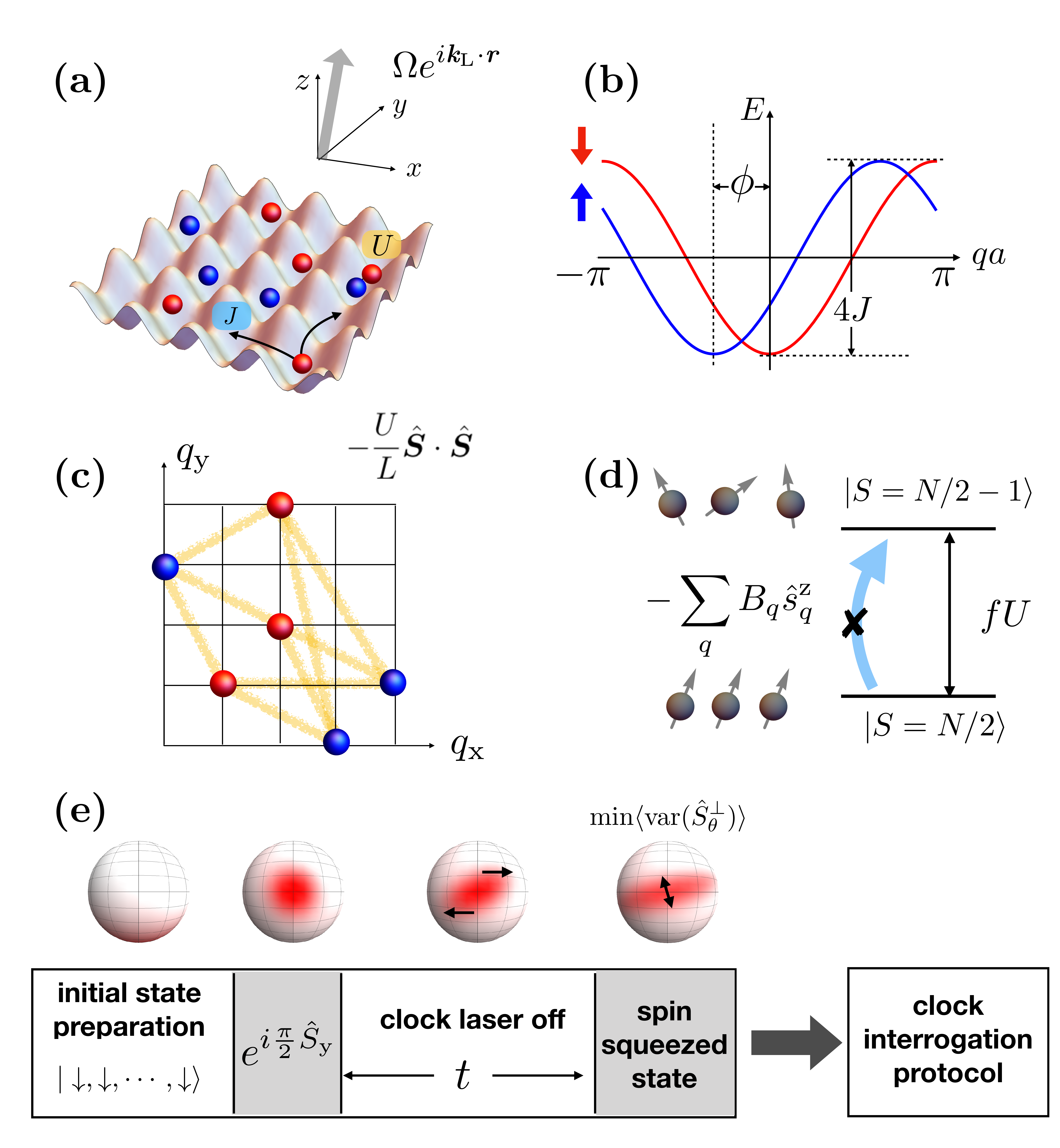}
\caption{{\bf Schematic of the setup for spin squeezing}.
({\bf a}) We consider $N$ fermionic atoms with two (pseudo-)spin components, represented by red and blue spheres, trapped in the ground band of an optical lattice (shown in 2D for the sake of presentation).
Atoms tunnel to neighboring sites at a rate $J$ and experience on-site interactions with strength $U$.
An external laser carrying a position dependent phase $e^{i\v k_{\t{L}}\cdot\v r}$ couples the spin states of the atoms.
({\bf b}) After a gauge transformation, different spin states exhibit different dispersion relations with a relative phase $\phi=k_{\text{L}} a$, where $a$ is the lattice spacing.
The external laser couples spin states with identical quasi-momenta $q$ in the gauge-transformed frame.
({\bf c}) If interactions are sufficiently weak, all motional degrees of freedom become frozen in momentum space, with atoms effectively pinned to fixed quasi-momentum modes $\v q$.
The dynamics on the frozen $\v q$-space lattice can then be mapped to a spin model in which collisional interactions correspond to a uniform, all-to-all ferromagnetic Heisenberg Hamiltonian with strength $U/L$, where $L$ is the total number of lattice sites.
({\bf d}) The spin dependence of the dispersion relation is captured by a mode-dependent axial field $B_q$ that generates inhomogeneous spin precession.
This axial field couples exchange-symmetric many-body Dicke states with total spin $S=N/2$ to spin-wave states with $S=N/2-1$.
The all-to-all interaction opens an energy gap $fU$ (with $f=N/L$ the filling fraction of spatial modes) between the Dicke states and the spin-wave states, which forbids population transfer between them in the weak-field limit.
({\bf e}) To generate spin squeezing via one-axis twisting, we initialize a product state with all spins polarized in $-z$ (i.e.~in $\ket{\dn}$), and apply a fast external laser pulse to rotate all spins into $x$.
We then let atoms freely evolve for a variable time $t$ (with a spin-echo pulse), after which the amount of spin squeezing can be determined experimentally from global spin measurements.
The spin-squeezed state can be used for a follow-up clock interrogation protocol (see Appendix \ref{sec:clock_interrogation}).
}
\label{fig:protocol_schematic}
\end{figure*}

We consider $N$ fermionic atoms with two spin states (labeled $\up$ and $\dn$) trapped in a 3D optical lattice.
In this discussion, these spin states are associated with the two electronic states of a nuclear-spin-polarized gas.
At sufficiently low temperatures, atoms  occupy the lowest Bloch band of the lattice and interact only through $s$-wave collisions.
A schematic of this system is provided in Fig.~\ref{fig:protocol_schematic}(a), where tight confinement prevents motion along the vertical direction ($z$), effectively forming a stack of independent 2D lattices.
For simplicity and without loss of generality, however, we first consider the case when tunneling can only occur along one direction, $x$, and thus model the system as living in  one dimension.

An external laser with Rabi frequency $\Omega$ and wavenumber $k_{\t{L}}$ along the tunneling axis resonantly couples atoms' internal states through the Hamiltonian $\hat H_{\t{laser}}/\hbar = \sum_j \Omega e^{ik_{\t{L}}x_j} \hat c_{j,\up}^\dag \hat c_{j,\dn} + \t{h.c.}$, where $\hat c_{j\alpha}$ is a fermionic annihilation operator for an atom on site $j$ with internal state $\alpha\in\set{\up,\dn}$ and $x_j$ is the position of site $j$.
This laser imprints a position-dependent phase that equates to a momentum kick $k_{\t{L}}$ when an atom changes internal states by absorbing or emitting a photon, thereby generating spin-orbit coupling\cite{wall2016synthetic, livi2016synthetic}.
After absorbing the position dependence of the laser Hamiltonian into fermionic operators through the gauge transformation $\hat c_{j,\up}\to e^{ik_{\t{L}} x_j}\hat c_{j,\up}$, which makes $\hat H_{\t{laser}}$ spatially homogeneous, the atoms are well-described in the tight-binding limit by the Fermi-Hubbard Hamiltonian\cite{esslinger2010fermihubbard}
\begin{align}
  \hat H_{\t{FH}}^{(\phi)}/\hbar
  &= -J \sum_j \p{e^{i\phi} \hat c_{j,\up}^\dag \hat c_{j+1,\up}
  + \hat c_{j,\dn}^\dag \hat c_{j+1,\dn} + \text{h.c.}} \notag \\
  &\qquad + U \sum_j \hat n_{j,\up} \hat n_{j,\dn},
\end{align}
where $J$ is the nearest-neighbor tunneling rate; the SOC angle $\phi\equiv k_{\t{L}}a$ determines the phase gained by spin-up atoms upon tunneling from site $j$ to site $j+1$ (in the gauge-transformed frame) with lattice spacing $a=x_{j+1}-x_j$; $U$ is the on-site interaction energy of two atoms; and $\hat n_{j\alpha}\equiv\hat c_{j\alpha}^\dag \hat c_{j\alpha}$ is a number operator.

The Fermi-Hubbard Hamiltonian can be re-written in the quasi-momentum basis with annihiliation operators $\hat c_{q\alpha}\equiv L^{-1/2}\sum_j e^{-iqx_j}\hat c_{j\alpha}$, where $q$ is a quasi-momentum and $L$ is the total number of lattice sites.
In this basis, the single-particle Hamiltonian exhibits shifted dispersion relations that signify spin-orbit coupling [see Fig.~\ref{fig:protocol_schematic}(b)]:
\begin{align}
  \hat H_{\t{FH,single}}^{(\phi)}/\hbar
  = -2J\sum_q\sp{\cos(qa + \phi) \hat n_{q,\up}
  + \cos(qa) \hat n_{q,\dn}}.
\label{eq:FermiHubbard}
\end{align}
When $U\lesssim J$, interaction energies are too weak for collisions to change the occupancies of single-particle quasi-momentum modes.
Atoms are then pinned to these modes, which form a lattice in quasi-momentum space [see Fig.~\ref{fig:protocol_schematic}(c)]\cite{bromley2018dynamics}.
In this strong-tunneling limit, the Fermi-Hubbard Hamiltonian [Eqn.~\eqref{eq:FermiHubbard}] can be mapped to a spin-$1/2$ system with a collective ferromagnetic Heisenberg interaction and an inhomogeneous axial field, given by\cite{martin2013quantum,rey2014probing, bromley2018dynamics}
\begin{align}
  \hat H_{\t{spin}}/\hbar
  = -\frac{U}{L} \hat{\v S} \c \hat{\v S}
  - \sum_q B_q \hat s_q^\z,
\label{eq:SpinModel}
\end{align}
where $\hat{\v S}=\sum_q\hat{\v s}_q$ is a collective spin operator; $\hat{\v s}_q$ is a spin-1/2 operator for mode $q$ with components $\hat{s}_q^{j=\x,\y,\z}\equiv\frac12\sum_{\alpha,\beta}\hat c_{q\alpha}^\dag\sigma^j_{\alpha\beta}\hat c_{q\beta}$ defined in terms of the Pauli matrices $\sigma^{j=\x,\y,\z}$; the sums over $q$ run over all occupied quasi-momentum modes; and $B_q \equiv -4J\sin(qa+\phi/2)\sin(\phi/2)$ is the SOC-induced axial field.

On its own, the collective Heisenberg term ($\sim\hat{\v S}\c\hat{\v S}$) in Eqn.~\eqref{eq:SpinModel} opens an energy gap $fU$, with $f\equiv N/L$ the filling fraction of spatial modes, between the collective Dicke states $\ket{S=N/2, M_S}$ and the remainder of Hilbert space\cite{rey2008manybody, martin2013quantum, norcia2018cavitymediated, smale2019observation} with $S<N/2$.
Here $S$ and $M_S$ respectively label the eigenvalues of the collective spin operators $\hat{\v S}\c\hat{\v S}$ and $\hat{S}_\z$, with eigenvalues $S(S+1)$ for non-negative $S\in\set{N/2,N/2-1,\cdots}$ and $M_S\in\set{-S,-S+1,\cdots,S}$.
The axial field $B_q$ generally couples states within the Dicke manifold to states outside it.
In the weak SOC limit (i.e.~$B_q\ll fU$), however, the interaction energy gap suppresses population transfer between states with different total spin $S$ [see Fig.~\ref{fig:protocol_schematic}(d)].
In this regime, the virtual occupation of states outside the Dicke manifold can be accounted for perturbatively.
The symmetries of SOC as expressed in Eqn.~\eqref{eq:SpinModel} dictate that this treatment should yield powers of $\hat S_\z$ when projected onto the collective Dicke manifold at higher orders in perturbation theory.
At second order in perturbation theory (see Appendix \ref{sec:derivation_OAT}), we thus find that SOC effectively yields a one-axis twisting (OAT) model widely known to generate squeezing dynamics\cite{kitagawa1993squeezed, ma2011quantum}:
\begin{align}
  \hat H_{\t{eff}}/\hbar
  = -\frac{U}{L} \hat{\v S} \c \hat{\v S}
  - \overline B \hat S_\z + \chi \hat S_\z^2,
  &&
  \chi \equiv \f{\widetilde B^2}{(N-1)fU},
  \label{eq:H_eff}
\end{align}
where $\overline B\equiv\sum_q B_q/N$ is the mean and $\widetilde B^2\equiv\sum_q\p{B_q-\overline B}^2/N$ the variance of the axial field.
The effect of the $\sim\hat{\v S}\c\hat{\v S}$ term is to generate a relative phase between states with different total spin $S$ and thus has no effect on dynamics  restricted to a fixed $S$.
Note also that  the collective spin rotation from $\overline B \hat S_\z$  can be eliminated by going into a rotating frame or by using a spin echo.

The entire protocol for preparing a squeezed state via OAT, sketched out in Fig.~\ref{fig:protocol_schematic}(e), reduces to a standard Ramsey protocol with a spin echo: after initially preparing a spin-down (i.e.~$-\uv z$) polarized sample of ultracold atoms populating the lowest Bloch band of a lattice, a fast $\pi/2$ pulse is applied with the clock laser to rotate all spin vectors into $+\uv x$.
The atoms then freely evolve for a variable time $t$ (possibly with spin-echo $\pi$-pulses), after which the amount of metrologically useful spin squeezing, measured by the Ramsey squeezing parameter
\begin{align}
  \xi^2 \equiv
  \min_\theta \braket{\t{var}(\hat S^\perp_\theta)}
  \times N/\abs*{\braket{\hat{\v S}}}^2,
  \label{eq:sqz}
\end{align}
can be determined experimentally from global spin measurements.
Here $\braket{\hat{\v S}}$ is the mean collective spin vector and $\braket{\t{var}(\hat S^\perp_\theta)}$ is the variance of spin measurements along an axis orthogonal to $\braket{\hat{\v S}}$, parameterized by the angle $\theta\in[0,2\pi)$.

The above protocol concerns only the preparation of a spin-squeezed state, which would then be used as an input state for a follow-up clock interrogation protocol without SOC.
While increasing the lattice depth to turn off SOC during clock interrogation is the simplest approach, this will limit the interrogation time due to light scattering (see discussion below).
Alternatively, it is possible to keep the same lattice depth used for the spin squeezing generation by adding a specific pulse sequence to suppress SOC.
See details in Appendix \ref{sec:clock_interrogation}.

\subsection{Model validity}
\label{sec:validity}

The validity of the OAT model in Eqn.~\eqref{eq:H_eff} relies on two key conditions concerning experimental parameter regimes.
First, the on-site interaction energy $U$ should not be much larger in magnitude than the tunneling rate $J$ (clarified below); otherwise, one cannot assume frozen motional degrees of freedom (i.e.~with atoms pinned to fixed quasi-momentum modes) and map the Fermi-Hubbard model to a spin model.
Second, the SOC-induced fields $B_q\sim J\sin(\phi/2)$ should be considerably smaller in magnitude than the interaction energy gap $fU$, as otherwise one cannot perturbatively transform SOC into OAT.
These two conditions can be satisfied by appropriate choices of $U/J$ and the SOC angle $\phi$, which are respectively controlled by tuning the lattice depth and changing the angle between the clock laser and the lattice axes [see Fig.~\ref{fig:protocol_schematic}(a)].

\begin{figure*}
\centering
\includegraphics[width=0.8\textwidth]{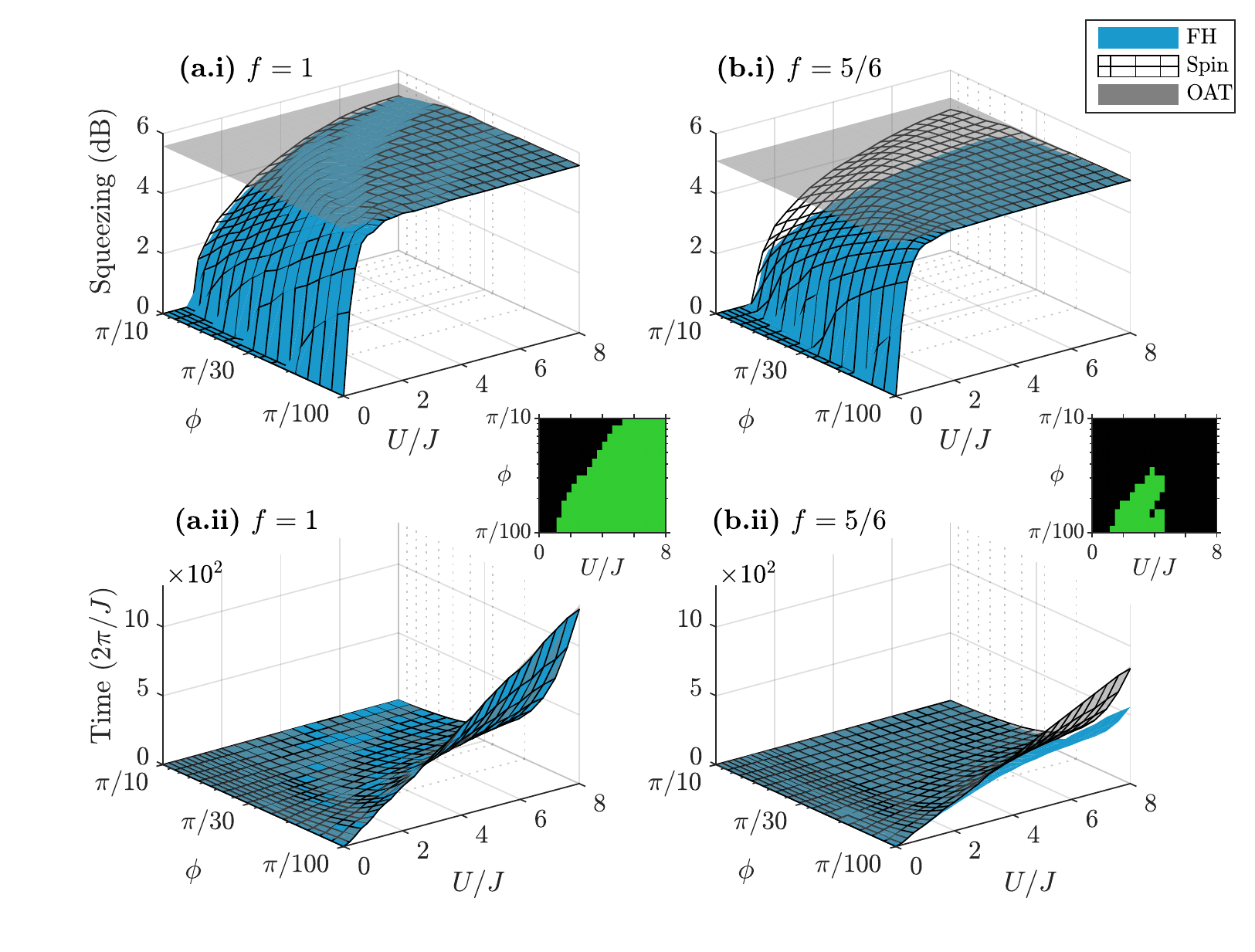}
\caption{{\bf Benchmarking the spin and one-axis twisting models}.
Comparisons of maximum squeezing (top panels, {\bf a.i} and {\bf b.i}) and optimal squeezing time (lower panels, {\bf a.ii} and {\bf b.ii}) between the Fermi-Hubbard (FH), spin, and one-axis twisting (OAT) models; obtained numerically via the protocol depicted in Fig.~\ref{fig:protocol_schematic}(e) in a 1D lattice with $L=12$ sites.
Results are shown for half filling with $N=12,f\equiv N/L=1$ (left panels, {\bf a.i} and {\bf a.ii}) and filling $f=5/6$ (right panels, {\bf b.i} and {\bf b.ii}) as a function of $U/J$ and the SOC angle $\phi$.
In both cases, the system is initialized in the corresponding ground state.
Insets for both $f=1$ and $f=5/6$ show (in green) regions of the $U$-$\phi$ plane in which both the optimal squeezing (in dB) and the corresponding squeezing time of all three models agree to within 20\%.
At half filling ({\bf a.i} and {\bf a.ii}), mode-changing collisions are suppressed by Pauli blocking, resulting in almost exact agreement between the FH and spin models; both of these models converge onto the OAT model in the gap-protected, weak SOC regime of large $U/J$ and small $\phi$.
The spin and OAT models show similar behavior away from half filling ({\bf b.i} and {\bf b.ii}), but the presence of mode-changing collisions results in their disagreement with the FH model as interactions begin to dominate at larger $U/J$.
Even below half filling, however, the FH exhibits comparable amounts of squeezing to the spin model across a broad range of $U/J$ and $\phi$, albeit at earlier times when $U/J\gtrsim2$.
}
\label{fig:model_benchmarking}
\end{figure*}

We demonstrate the importance of these conditions in Fig.~\ref{fig:model_benchmarking}, where we show numerical results from exact simulations of a 1D system with $L=12$ sites.
Therein, optimal squeezing achievable under unitary dynamics is provided in dB, i.e.~$-10\log_{10}(\xi_{\t{opt}}^2)$, while the time at which this squeezing occurs is provided in units of the nearest-neighbor tunneling time $2\pi/J$.
At $f=1$ atom per lattice site, i.e.~half filling of all atomic states in the lowest Bloch band, the spin model [Eqn.~\eqref{eq:SpinModel}] agrees almost exactly with the Fermi-Hubbard (FH) model [Eqn.~\eqref{eq:FermiHubbard}] up through (and exceeding) $U/J=8$. The agreement at half filling ($f=1$) is assisted by Pauli blocking of mode-changing collisions.
Below half filling ($f=5/6$), these two (FH and spin) models show good agreement at $U/J\lesssim2$, while at $U/J\gtrsim2$ mode-changing collisions start to become relevant and invalidate the frozen-mode assumption of the spin model.
Note that we chose filling $f=5/6$ to demonstrate that our protocol should work, albeit sub-optimally, even in this highly hole-doped case; in practice, optimized experiments are capable of achieving fillings closer to the optimal $f=1$\cite{brown2017spinimbalance}.
Interestingly, even with mode-changing collisions the Fermi-Hubbard model exhibits comparable amounts of squeezing to the spin model, and achieves this squeezing in less time.
The spin and OAT models agree in the regime of weak SOC with $\widetilde{B}\sim J\sin(\phi/2)\ll fU$, and exhibit different squeezing behaviors outside this regime as single-particle spin dephasing can no longer be treated as a weak perturbation to the spin-locking  interactions.

In realistic implementations, the Gaussian profile of the laser beams always introduces an additional effective harmonic potential that modifies the translational invariance assumed so far.
We present a detailed discussion of the role of the harmonic trap in Appendix \ref{sec:harmonic_trap}, where we demonstrate that the addition of harmonic confinement barely modifies the achievable spin squeezing with currently accessible trapping frequencies.
We find that the existence of single-particle localized modes in the lattice with harmonic confinement\cite{rey2005ultracold, pupillo2006extended} helps to protect spin squeezing and shifts the optimal parameter window to $U/J\gtrsim 2$.

\subsection{Two-axis twisting}
\label{sec:TAT}

The above scheme for OAT achieves optimal spin squeezing that scales as $\xi^2_{\t{opt}}\sim N^{-2/3}$ with minimal intervention, i.e.~a standard Ramsey protocol.
Further improvements upon this scheme can be made by introducing a time-dependent driving field that transforms the OAT Hamiltonian into a two-axis twisting (TAT) one.
While the OAT model initially generates squeezing faster than the TAT model, the squeezing generation rate of OAT (measured in dB per second) falls off with time, while the squeezing generation rate for TAT remains approximately constant until reaching Heisenberg-limited amount of spin squeezing with $\xi^2_{\t{opt}}\sim N^{-1}$\cite{kitagawa1993squeezed}.

There are two general strategies for converting OAT into TAT: by use of either a pulsed\cite{liu2011spin} or continuous\cite{huang2015twoaxis} drive protocol.
For simplicity, we consider the latter in this  work, although the pulsed protocol could provide additional advantages, as explained at the end of Appendix \ref{sec:clock_interrogation}.
Following the prescription in Ref.~[\citenum{huang2015twoaxis}], we use the clock laser to apply an amplitude-modulated drive $\hat H_{\t{drive}}(t)/\hbar=\Omega_0\cos(\omega t)\hat{S}_\x$.
If the modulation frequency $\omega$ satisfies $\omega\gg N\chi$ and $\J_0\p{2\Omega_0/\omega}=\pm1/3$, where $\chi$ is the OAT squeezing strength in Eqn.~\eqref{eq:H_eff} and $\J_0$ is the zero-order Bessel function of the first kind, then up to (i) an $\sim\hat{\v S}\c\hat{\v S}$ term that contributes only overall phase factors, and (ii) an $\sim\hat S_\z$ term that can be eliminated with a simple dynamical decoupling pulse sequence (see Appendix \ref{sec:dynamical_decoupling}), the effective Hamiltonian becomes $\hat H_{\t{TAT}}^{(+)}/\hbar = (\chi/3)(\hat S_\z^2-\hat S_\x^2)$ or $\hat H_{\t{TAT}}^{(-)}/\hbar = (\chi/3)(\hat S_\y^2-\hat S_\x^2)$ (see Appendix \ref{sec:derivation_TAT}), which squeezes an initial state polarized along the $y$ or $z$ axis, respectively.

\section{Experimental implementation and practical considerations}

Thus far, we have largely considered the general preparation of spin-squeezed states with the Fermi-Hubbard model.
Here, we discuss the specific implementation of the above protocols in the state-of-the-art 3D $^{87}$Sr optical lattice clock (OLC).
If successful, such an implementation would (to our knowledge) for the first time break through the proof-of-principle stage of spin squeezing efforts, and achieve a genuine metrological enhancement of a world-class quantum sensor.

As required for our protocol, 3D $^{87}$Sr OLC has demonstrated the capability to load a quantum degenerate gas into a 3D lattice at the ``magic wavelength'' ($\lambda_{\t{lattice}}=2a\approx813$ nm) for which both the ground ($^1S_0,\dn$) and first excited ($^3P_0,\up$) electronic states (i.e.~the ``clock states'') of the atoms experience the same optical potential\cite{campbell2017fermidegenerate}.
Furthermore, the 3D $^{87}$Sr OLC currently operates at sufficiently low temperatures to ensure vanishing population above the lowest Bloch band, such that its dynamics are governed by the Fermi-Hubbard Hamiltonian [Eqn.~\eqref{eq:FermiHubbard}]\cite{esslinger2010fermihubbard}.

An external clock laser with wavelength $\lambda_{\t{L}}\approx698$ nm resonantly interrogates the $^1S_0$ and $^3P_0$ states of the atoms and generates spin-orbit coupling (SOC)\cite{wall2016synthetic}.
While the relative wavelengths of the lattice and clock lasers do not allow for weak SOC along all three lattice axes, weak SOC along two axes can be implemented by, for example, (i) fixing a large lattice depth along the $z$ axis, effectively freezing atomic motion along  $z$, and then (ii) making the clock laser nearly collinear with the $z$ axis, with only a small projection of its wavenumber $\v k_{\t{L}}$ onto the $x$-$y$ plane [see Figure \ref{fig:protocol_schematic}(a)].
The entire 3D OLC then factorizes into an array of independent 2D systems with $N=f\ell^2$ atoms each, where $\ell$ is the number of lattice sites along each axis of the lattice.
As in the 1D case, atoms within the 2D system experience all-to-all interactions, as well as spin-orbit coupling along two directions characterized by SOC angles $\phi_{\x,\y}=k_{\t{L}}^{\x,\y}a$.
Generally speaking, higher-dimensional systems (e.g.~2D vs.~1D) are more desirable because they allow packing more interacting atoms into a fixed system volume, thereby increasing the maximally attainable amount of spin squeezing.

\begin{figure*}
\centering
\includegraphics{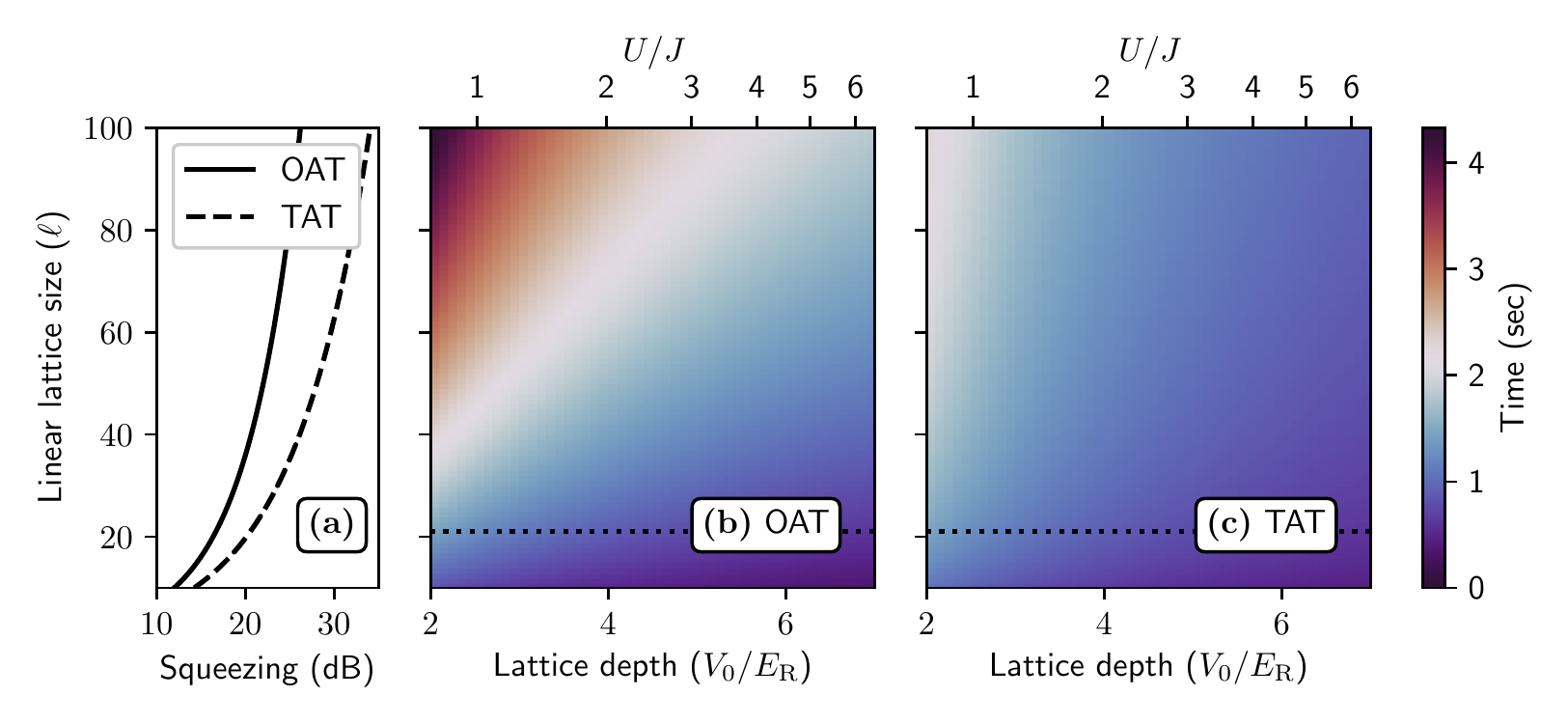}
\caption{{\bf Optimal squeezing} with one- and two-axis twisting in a 2D section of the 3D $^{87}$Sr optical lattice clock.
({\bf a}) The maximum amount of squeezing depends only on the atom number $N=\ell^2$, where $\ell$ is the number of lattice sites along each axis of the lattice.
While the time scales for squeezing generally depend on several experimental parameters, the time at which maximal squeezing occurs can be minimized at any given lattice depth $V_0$ by choosing SOC angles $\phi$ that saturate $\widetilde{B}/U\approx0.05$, where $\widetilde{B}$ is the variance of the SOC-induced axial field and $U$ is the two-atom on-site interaction energy.
Panels ({\bf b}, {\bf c}) show these minimal squeezing times as a function of the depth $V_0$ and linear size $\ell$ of the lattice.
Lattice depths $V_0$ are normalized to the atomic lattice recoil energy $E_{\t{R}}$, and the upper axis on panels ({\bf b}, {\bf c}) marks values of  $U/J$ at fixed lattice depths.
In general, TAT achieves more squeezing than OAT for any system size, and achieves optimal squeezing faster for $N\gtrsim400$ atoms, as denoted by a dotted line in panels ({\bf b}, {\bf c}).
}
\label{fig:optima_coherent}
\end{figure*}

Figure \ref{fig:optima_coherent} shows, for both OAT and TAT protocols, the maximally attainable amount of spin squeezing and the shortest time at which it occurs as a function of the lattice depth $V_0$ and linear lattice size $\ell$ in a single half-filled 2D layer (i.e.~$f=1,N=\ell^2$) of the 3D OLC.
Atoms are confined along the direction transverse to the 2D layer by a lattice of depth 60 $E_{\t{R}}$, where $E_{\t{R}}$ is the atomic lattice recoil energy.
The maximally attainable amount of spin squeezing by each protocol in Fig.~\ref{fig:optima_coherent} depends only on the atom number $N$, while the shortest attainable time is determined by choosing the largest SOC angles $\phi_\x=\phi_\y\equiv\phi$ which saturate $\widetilde{B}/U\approx0.05$.
We impose this constraint on $\widetilde{B}/U$ to ensure validity of the OAT Hamiltonian perturbatively derived in Appendix \ref{sec:derivation_OAT} (see also Appendix \ref{sec:benchmarking}).

\begin{figure*}
\centering
\includegraphics{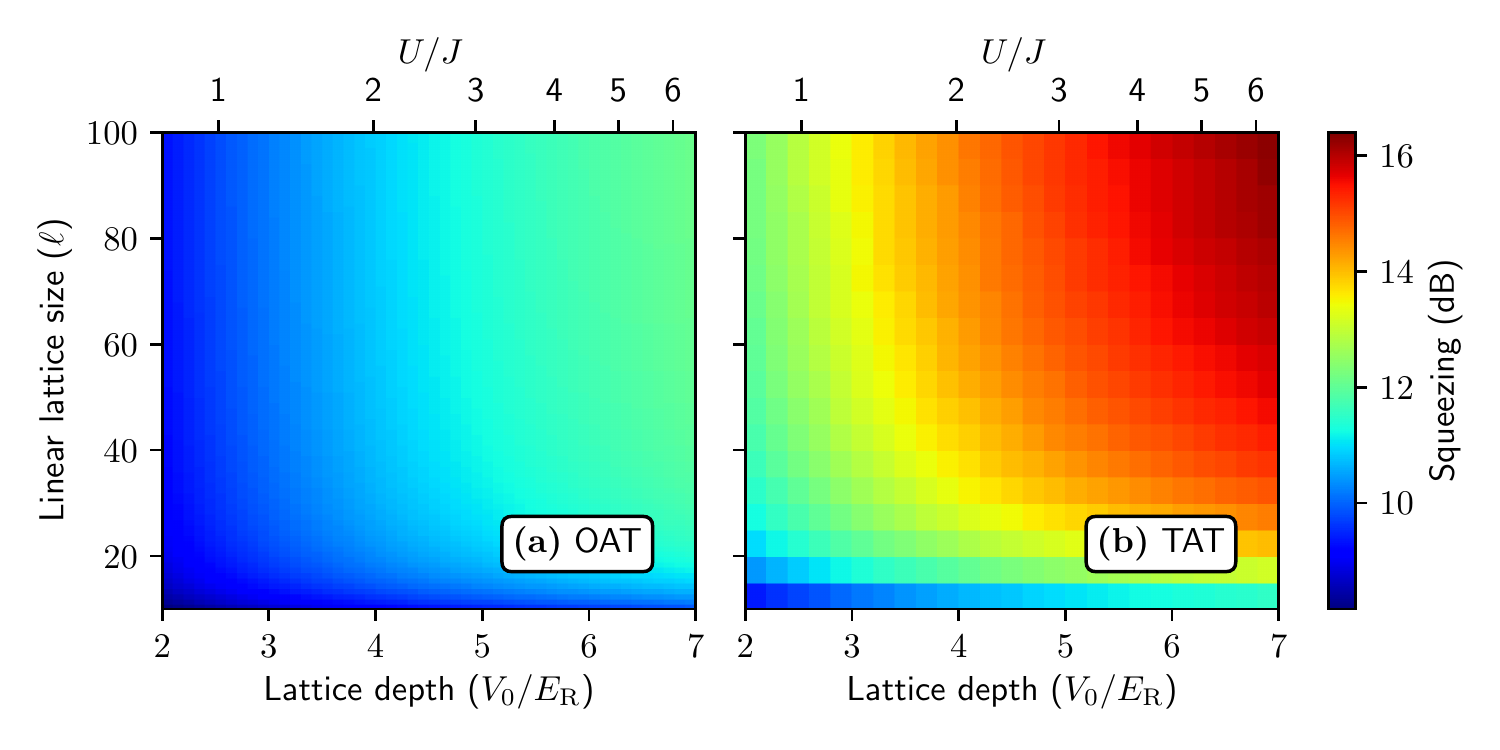}
\caption{{\bf Optimal squeezing with decoherence} via one- and two-axis twisting in a 2D section of the 3D $^{87}$Sr optical lattice clock (OLC).
In practice, decoherence due to light scattering limits the amount of squeezing that is attainable in the the 3D $^{87}$Sr OLC.
Due to growing squeezing times with increasing system size, the maximal squeezing obtainable via OAT saturates past $\ell\approx30$ sites along each axis of the lattice, with $N\approx10^3$ atoms total.
The more favorable size-dependence of TAT time scales,  however, allow for continued squeezing gains through $\ell=100$ ($N=10^4$).
While the OAT results in ({\bf a}) are exact, the TAT results in ({\bf b}) reflect only a lower bound on the maximum squeezing obtainable, albeit one that is likely close (within a few dB) to the actual value.
Optimal squeezing times in the presence of decoherence are generally smaller than the corresponding times shown in Fig.~\ref{fig:optima_coherent}, as decoherence typically degrades squeezing before it reaches the decoherence-free maximum.
The decoherence considered in this work also limits maximally achievable squeezing to $\sim20$ dB less than the decoherence-free maxima shown in Fig.~\ref{fig:optima_coherent}.
Sample plots of squeezing over time for particular choices of lattice size ($\ell$) and depth ($V_0/E_{\t{R}}$) are provided in Appendix \ref{sec:time_series}.
}
\label{fig:optima_decoherence}
\end{figure*}

Currently, light scattering from the lattice beams induces decoherence of the clock on a time scale of $\sim10$ seconds\cite{goban2018emergence, hutson2019engineering}, which is much shorter than the natural $^3P_0$ lifetime of $\sim160$ seconds (see Appendix \ref{sec:decoherence}).
This limitation imposes significant constraints on achievable spin squeezing, as shown in Figure \ref{fig:optima_decoherence} where the maximal squeezing with spin decay in the OAT case was determined using exact expressions for spin correlators derived in Ref.~[\citenum{foss-feig2013nonequilibrium}], while in the TAT case these correlators were determined by solving Heisenberg equations of motion for collective spin operators\cite{perlin2019shorttime} (see Appendix \ref{sec:collective_simulation}).
Due to the fast growth of Heisenberg operators in systems with all-to-all interactions, the latter method is not always capable of simulating up to the optimal squeezing time, and thus only provides a lower bound on the maximal squeezing theoretically obtainable via TAT.

The results in Fig.~\ref{fig:optima_decoherence} show that squeezing via OAT saturates with system size around $N\approx10^3$ ($\ell\approx30$), while TAT allows for continued squeezing gains through $N=10^4$ ($\ell=100$).
Even with decoherence, our protocol may realistically generate $\sim10$--$14$ dB of spin squeezing in $\sim1$ second with $\sim10^2$--$10^4$ atoms in a 2D section of the lattice, which is compatible with the atom numbers and interrogation times of state-of-the-art optical lattice clocks\cite{campbell2017fermidegenerate, marti2018imaging}.
This amount of spin squeezing exceeds those reported in the ground-state nuclear spin sublevels of a state-of-the-art ${}^{171}$Yb OLC ($\sim6.5$ dB)\cite{braverman2019nearunitary}.
While the latter protocol might be used to transfer spin squeezing to the electronic clock state, to date there has been no demonstration of spin squeezing in an optical clock transition.

In addition to light scattering, $p$-wave losses from inelastic ${}^3 P_0$ collisions\cite{martin2013quantum, zhang2014spectroscopic, bishof2011inelastic} can also degrade the maximum achievable spin squeezing, which becomes more pronounced for shallower lattices.
More details on $p$-wave losses are discussed in Appendix \ref{sec:inelastic_collision}, where we show that operating at lattice depths $V_0\gtrsim 7 E_{\t{R}}$ may be necessary to suppress the impact of inelastic collisions on spin squeezing, at the cost of slightly increasing light scattering.

The sources of decoherence considered above are not fundamental, and can be avoided by, for instance, using two nuclear spin levels as spin-1/2 degrees of freedom that are interrogated by far-detuned Raman transitions instead of a direct optical transition\cite{mancini2015observation}.
The strength of SOC for Raman transitions is tunable and, moreover, the lifetimes of ground nuclear spin levels are longer than 100 seconds in the lattice\cite{goban2018emergence}.
In this case, our protocol for preparing a squeezed state would additionally end with a coherent state transfer from nuclear to electronic degrees of freedom to retain metrological utility for the atomic clock.
If, for example, the $-9/2$ and $-7/2$ nuclear spin states are used for the preparation of a squeezed state, then the collective-spin entanglement of atoms can be transferred to electronic states at the end of the spin squeezing protocol with a $\sigma^-$ polarized $\pi$ pulse.
Such a pulse can transfer $\ket{g,-7/2}$ to $\ket{e,-9/2}$ without affecting $\ket{g,-9/2}$, where $g$ and $e$ respectively denote the ground and excited (electronic) clock states.

\section{Conclusions}

We have proposed a new protocol to generate spin squeezing in a fermionic 3D optical lattice clock by combining nominally undesirable atomic collisions with spin-orbit coupling.
To our knowledge, this is the first proposal to use quantum correlations in a many-body fermionic system to push state-of-the-art quantum sensors beyond the independent-particle regime, thereby achieving a genuine quantum advantage.
Such capability could allow for major improvements in clock sensitivity and bandwidth, enhancing not only traditional timekeeping applications such as measurement standards, navigation (GPS), and telecommunications, but also geodesy and gravitational wave detection, precision tests of fundamental physics, and the search for new physics beyond the standard model\cite{safronova2018search}.

\section{Acknowledgments}

We acknowledge helpful discussions with M.~Norcia, C.~Sanner, and M.~Mamaev.
This work is supported by the Air Force Office of Scientific Research (AFOSR) grant FA9550-18-1-0319; the AFOSR Multidisciplinary University Research Initiative (MURI) grant; the Defense Advanced Research Projects Agency (DARPA) and Army Research Office (ARO) grant W911NF-16-1-0576; the National Science Foundation (NSF) grant PHY-1820885; JILA-NSF grant PFC-1734006; and the National Institute of Standards and Technology (NIST).

\appendix

\renewcommand\thefigure{\thesection\arabic{figure}}

\section{Derivation of the effective one-axis-twisting model}
\label{sec:derivation_OAT}

Suppose we have a Hamiltonian of the form ($\hbar=1$)
\begin{align}
  H = H_0 + V,
\end{align}
with
\begin{align}
  H_0 = - \f{U}{L} \v S\c\v S,
  &&
  V = - \sum_n B_n s_\z^{(n)} + \Omega S_\x,
\end{align}
and we consider $N$-particle states initially in the ground-state manifold $\G_0$ of $H_0$, which have total spin $S=N/2$.
If the largest eigenvalue of $V$ is smaller in magnitude than half of the collective spin gap $NU/L=fU$, i.e.~the energy gap under $H_0$ between $\G_0$ and its orthogonal complement $\E_0$, then we can formally develop a perturbative treatment for the action of $V$ on $\G_0$.
Such a treatment yields an effective Hamiltonian on $\G_0$ of the form $H_{\t{eff}}=\sum_pH_{\t{eff}}^{(p)}$, where $H_{\t{eff}}^{(p)}$ is order $p$ in $V$.
Letting $\P_0$ ($\Q_0$) be a projector onto $\G_0$ ($\E_0$), we define the super-operators $\O$ and $\L$ by
\begin{align}
  \O V &\equiv \P_0 V \Q_0 + \Q_0 V \P_0, \\
  \L V &\equiv \sum_{\alpha,\beta}
  \f{\op{\alpha}\O V \op{\beta}}{E_\alpha-E_\beta},
\end{align}
where $H_0 = \sum_\alpha E_\alpha \op\alpha$.
The first few terms in the expansion of the effective Hamiltonian $H_{\t{eff}}$ are then, as derived in Ref.~[\citenum{bravyi2011schrieffer}],
\begin{align}
  H_{\t{eff}}^{(0)} = \P_0 H_0 \P_0,
  &&
  H_{\t{eff}}^{(1)} = \P_0 V \P_0,
\end{align}
and
\begin{align}
  H_{\t{eff}}^{(2)} = -\f12 \P_0 \sp{\O V,\L V}_- \P_0,
\end{align}
with $\sp{X,Y}_-\equiv XY-YX$.
The zero-order effective Hamiltonian $H_{\t{eff}}^{(0)}=H_0$ within the ground-state manifold.
To calculate $H_{\t{eff}}^{(1)}$, we note that the ground-state manifold $\G_0$ is spanned by the Dicke states
\begin{align}
  \ket{m} \propto S_+^{N/2+m} \ket{\downarrow}^{\otimes n},
  &&
  S_+ \equiv \sum_n s_+^{(n)},
\end{align}
in terms of which we can expand the collective spin-z operator as $S_\z=\sum_m m\op{m}$.
We can likewise expand the collective spin-x operator $S_\x$ in terms of $x$-oriented Dicke states $\ket{m_\x}$ as $S_\x=\sum_m m\op{m_\x}$.
The ground-state projector $\P_0$ onto $\G_0$ can be expanded in either basis as $\P_0=\sum_m\op{m}=\sum_m\op{m_\x}$.
Defining the mean and residual fields
\begin{align}
  \overline B \equiv \f1N \sum_n B_n,
  &&
  b_n \equiv B_n - \overline B,
\end{align}
we can then write
\begin{align}
  V &= -\sum_n \p{b_n+\overline B} s_\z^{(n)} + \Omega S_\x \notag \\
  &= -\sum_n b_n s_\z^{(n)} - \overline B S_\z + \Omega S_\x,
\end{align}
and in turn
\begin{align}
  H_{\t{eff}}^{(1)}
  &= \P_0\p{-\sum_n b_n s_\z^{(n)} - \overline B S_\z + \Omega S_\x} \P_0 \notag \\
  &= -\sum_n b_n \P_0 s_\z^{(n)} \P_0 - \overline B S_\z + \Omega S_\x,
\end{align}
where we used the fact that $\P_0 S_{j=\z,\x} \P_0 = S_j$ within the ground-state manifold.
By construction, the residual fields are mean-zero, i.e.~$\sum_n b_n=0$.
Using the particle-exchange symmetry of the Dicke states, we can therefore expand
\begin{multline}
  \sum_n b_n \P_0 s_\z^{(n)} \P_0
  = \sum_{n,m,m'} b_n \op{m} s_\z^{(n)} \op{m'} \\
  = \sum_n b_n \sum_{m,m'} \op{m} s_\z^{(1)} \op{m'}
  = 0,
\end{multline}
which implies
\begin{align}
  H_{\t{eff}}^{(1)} = - \overline B S_\z + \Omega S_\x.
\end{align}
To calculate the second-order effective Hamiltonian $H_{\t{eff}}^{(2)}$, we let $\B_0\p{\E_0}$ denote an eigenbasis of $H_0$ for the excited subspace $\E_0$, and set the ground-state energy to 0.
We then define the operator
\begin{align}
  \I \equiv \sum_{\ket\alpha\in\B_0\p{\E_0}} \f{\op\alpha}{E_\alpha},
\end{align}
which sums over projections onto excited states with corresponding energetic suppression factors, in terms of which we can write
\begin{align}
  H_{\t{eff}}^{(2)} = -\P_0 V \I V \P_0,
\end{align}
which is simply an operator-level version of the textbook expression for second-order perturbation theory.
The only part of $V$ which is off-diagonal with respect to the ground- and excited-state manifolds $\G_0$ and $\E_0$ is $-\sum_n b_n s_\z^{(n)}$, and the individual spin operators in this remainder can only change the total spin $S$ by at most 1.
It is therefore sufficient to expand $\I$ in a basis for states which span the image of $\G_0$ under all $s_\z^{(n)}$ within the $S=N/2-1$ manifold.
Such a basis is provided by the spin-wave states
\begin{align}
  \ket{mk}
  \propto
  \sum_{n=1}^N e^{2\pi ikn/N} s_+^{(n)} \ket{m-1},
\end{align}
for $k=1,2,\cdots,N-1$\cite{swallows2011suppression}.
Using the fact that all spin-$z$ operators preserve the projection of total spin onto the $z$ axis, we then have that
\begin{multline}
  H_{\t{eff}}^{(2)}
  = -\f1{fU} \sum_{m,k,n,n'} b_n b_{n'}
  \op{m} s_\z^{(n)} \ket{mk} \\
  \times \bra{mk} s_\z^{(n')} \op{m},
  \label{eq:general_H_eff_2}
\end{multline}
where the relevant matrix elements between the Dicke states and the spin-wave states are\cite{swallows2011suppression}
\begin{align}
  \bk{m|s_\z^{(n)}|mk}
  = e^{2\pi i k n/N} \sqrt{\f{(N/2)^2-m^2}{N^2 (N-1)}},
\end{align}
which implies
\begin{multline}
  H_{\t{eff}}^{(2)}
  = -\f1{fU} \sum_m \f{(N/2)^2-m^2}{N^2 (N-1)} \op{m} \\
  \times \sum_{k,n,n'} b_n b_{n'} e^{2\pi ik\p{n-n'}/N}.
\end{multline}
Using the fact that $\sum_nb_n=0$, we can expand
\begin{align}
  \sum_{k,n,n'} b_n b_{n'} e^{2\pi ik\p{n-n'}/N}
  &= \sum_{n,n'} b_n b_{n'} \sum_{k=1}^{N-1} e^{2\pi ik\p{n-n'}/N} \notag \\
  &= \sum_{n,n'} b_n b_{n'} \sum_{k=0}^{N-1} e^{2\pi ik\p{n-n'}/N},
\end{align}
where the sum over $k$ vanishes for $n\ne n'$ and equals $N$ when $n=n'$, so
\begin{align}
  \sum_{k,n,n'} b_n b_{n'} e^{2\pi ik\p{n-n'}/N}
  = N^2 \widetilde B^2,
\end{align}
where
\begin{align}
  \widetilde B^2
  \equiv \f1N \sum_n b_n^2 = \f1N \sum_n \p{B_n - \overline B}^2.
  \label{eq:sum_knn}
\end{align}
We therefore have that
\begin{align}
  H_{\t{eff}}^{(2)}
  = -\sum_m \f{\p{N/2}^2-m^2}{\p{N-1}fU}~ \widetilde B^2 \op{m},
\end{align}
where the $\p{N/2}^2$ term contributes a global energy shift which we can neglect, while the $m^2$ term is proportional to $m^2\op{m}=S_\z^2$.
In total, the effective Hamiltonian through second order in perturbation theory is thus
\begin{align}
  H_{\t{eff}}
  = -\f{U}{L}\v S\c\v S - \overline B S_\z + \Omega S_\x + \chi S_\z^2,
\label{eq:H_OAT}
\end{align}
with
\begin{align}
  \chi \equiv \f{\widetilde B^2}{\p{N-1}fU}.
\end{align}
We benchmark the validity of this effective Hamiltonian via exact simulations of the spin [Eqn.~\eqref{eq:SpinModel}] and OAT [Eqn.~\eqref{eq:H_eff}] Hamiltonians in a system of 20 spins, finding that the relative error in maximal squeezing (in dB) of the OAT model is less than 3\% when $\widetilde{B}/U<0.06$ (see Appendix \ref{sec:benchmarking}).

\section{Numerical benchmarking of the OAT model}
\label{sec:benchmarking}
\setcounter{figure}{0}

\begin{figure}
\centering
\includegraphics[width=0.35\textwidth]{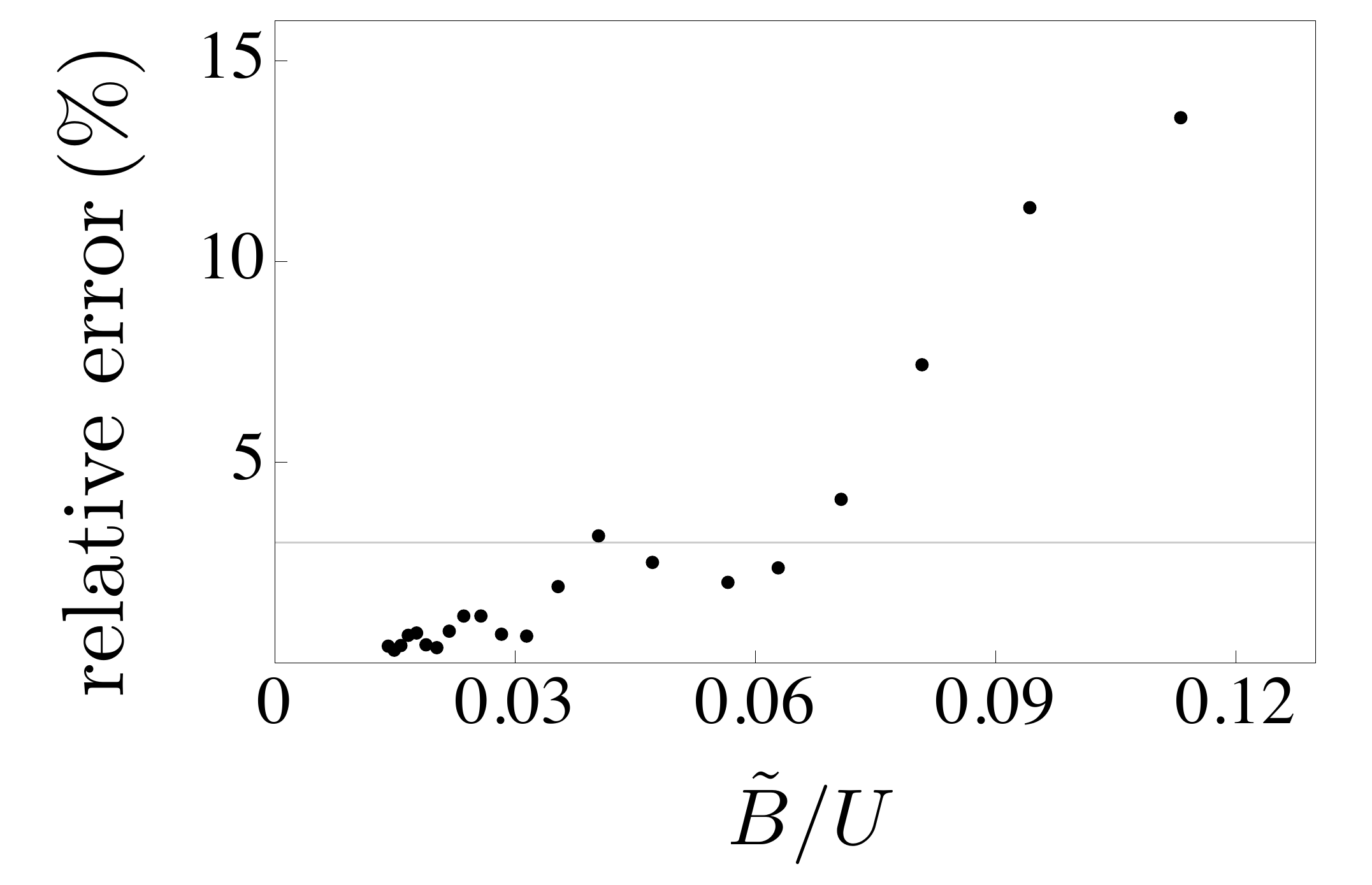}
\caption{{\bf Relative error} between maximal squeezing (measured in dB) obtained by the OAT [Eqn.~\eqref{eq:H_eff}] and spin [Eqn.~\eqref{eq:SpinModel}] models of the main text in a system of 20 particles.
The OAT model correctly captures the maximal squeezing (in dB) of the spin model to within 3\% (marked by the horizontal reference line) within the gap-protected regime $\widetilde{B}/U<0.06$.
}
\label{fig:relative_error}
\end{figure}

Here we provide additional information about our benchmarking of the one-axis twisting model against the spin model.
This benchmarking was performed via exact simulations of a 20-spin system.
Fig.~\ref{fig:relative_error} shows the relative error in maximal squeezing of the OAT model (measured against the spin model) as a function of the reduced field variance $\widetilde{B}/U$.
Here squeezing is measured in decibels (dB) by $-10\log_{10}\xi^2$ for the squeezing parameter $\xi^2$ define in Eqn.~\eqref{eq:sqz}.
The relative error in maximal squeezing (in dB) by the OAT model is less than 3\% when $\widetilde{B}/U<0.06$.

\begin{figure*}
\centering
\includegraphics[width=0.75\textwidth]{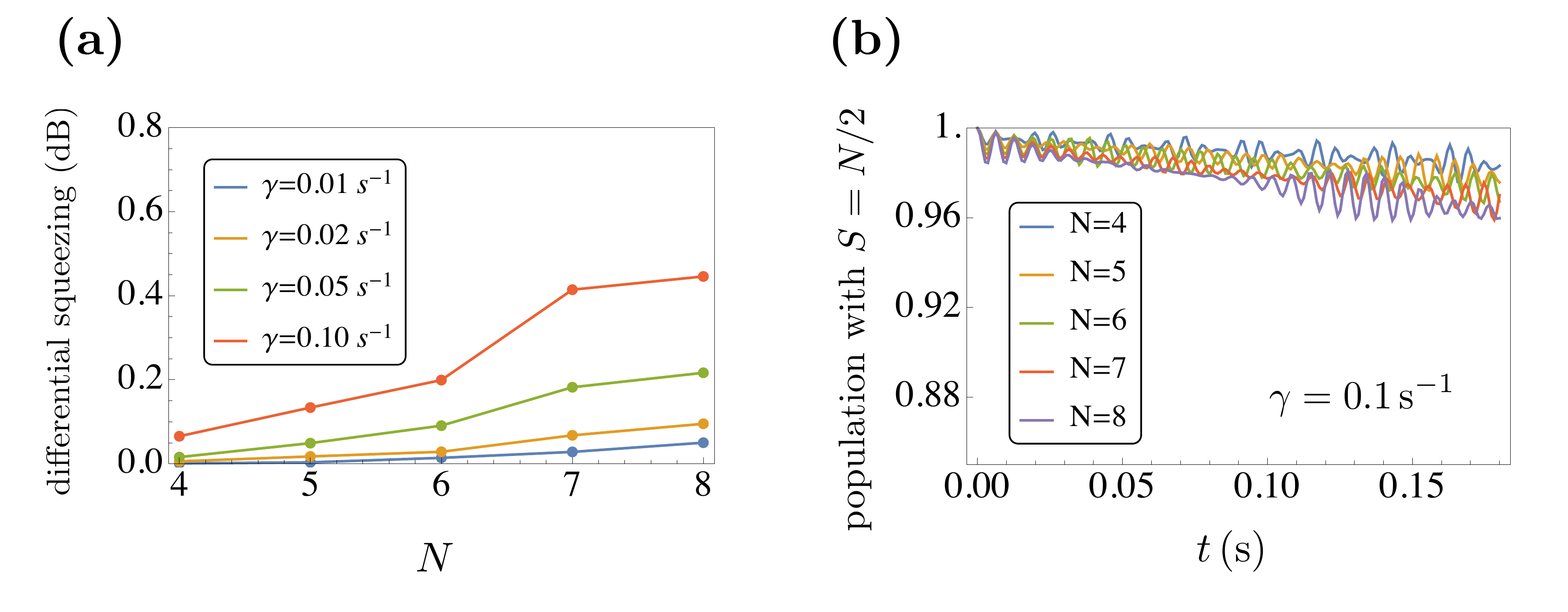}
\caption{{\bf Comparison between the OAT and the spin model in the presence of decoherence.}
({\bf a}) The difference between the maximal squeezing (measured in dB) obtained by the OAT [Eqn.~\eqref{eq:H_eff}] and spin [Eqn.~\eqref{eq:SpinModel}] models increases with the particle number $N$ and the single-particle spontaneous emission rate $\gamma$.
This disagreement is attributed in part to the fact that spontaneous emission transfers population of the collective spin state outside of the Dicke manifold, violating an assumption of the OAT model; see panel ({\bf b}).
The rate of population transfer outside of the Dicke manifold increases with both particle number and spontaneous emission rate.
(Parameters for simulations in this figure: $U=1000$ Hz, $J=200$ Hz, and $\phi=\pi/20$).
}
\label{fig:benchmarking_decay}
\end{figure*}

In principle, spin-changing decoherence compromises the validity of the OAT model, as its perturbative derivation in Appendix \ref{sec:derivation_OAT} relies on spin population remaining primarily within the Dicke manifold.
This assumption breaks down in the presence of, for example, spontaneous emission, which transfers population outside of the Dicke manifold.
Nonetheless, we find decent agreement between the OAT and spin models when decoherence is sufficiently weak (see Fig.~\ref{fig:benchmarking_decay}).

\section{Two-axis twisting, decoherence, and the residual axial field}
\label{sec:derivation_TAT}

The protocol we use to transform one-axis twisting (OAT) into two-axis twisting (TAT) is as previously proposed in Ref.~[\citenum{huang2015twoaxis}]; we provide a summary of this protocol here, in addition to some brief discussion of its implications for decoherence and the residual $\sim\v S\v\c\v S$ and $\sim S_\z$ terms of our OAT protocol.
The TAT protocol begins with the OAT Hamiltonian with a time-dependent transverse field,
\begin{align}
  H = \chi S_\z^2 + \Omega\p{t} S_\x,
  &&
  \Omega\p{t} = \beta\omega \cos\p{\omega t},
  \label{eq:driven_OAT}
\end{align}
where $\beta$ is the modulation index of the driving field and the drive frequency $\omega\gg N\chi$, with $N$ the total number of spins.
Moving into the rotating frame of $\Omega\p{t}S_\x$ subtracts this term from the Hamiltonian, and transforms operators $\O$ as
\begin{align}
  \O \to U\p{t}^\dag \O U\p{t},
\end{align}
where
\begin{align}
  U\p{t} \equiv \exp\sp{-i\int_0^t\d\tau~\Omega\p{\tau}S_\x}
  = \exp\sp{-i\beta\sin\p{\omega t}S_\x}.
\end{align}
In particular, the operators $\tilde S_\pm\equiv-S_\z\pm iS_\y$ (i.e.~the raising and lowering operators in the $x$ basis) transform simply as
\begin{align}
  \tilde S_\pm
  \to U^\dag \tilde S_\pm U
  = e^{\pm i\beta\sin\p{\omega t}} \tilde S_\pm.
\end{align}
For any operator $\O$ and drive frequency $\omega\gg\norm{\O}$, where $\norm{\O}\equiv\max_\psi\sqrt{\braket{\psi|\O^\dag\O|\psi}}$ is the operator norm of $\O$ (i.e.~the magnitude of the largest eigenvalue of $\O$), we can generally make a secular approximation to say
\begin{align}
  e^{\pm im\beta\sin\p{\omega t}} \O
  &= \sum_{n=-\infty}^\infty \J_n\p{\pm m\beta} e^{in\omega t} \O \notag \\
  &\approx \J_0\p{\pm m\beta} \O
  = \J_0\p{m\beta} \O,
\end{align}
where $\J_n$ is the $n$-th order Bessel function of the first kind.
Expanding $S_\z^2=\f14\p{\tilde S_+ + \tilde S_-}^2$, one can thus work out that the effective Hamiltonian in the rotating frame of the drive is
\begin{align}
  H_{\t{eff}}
  \approx \f{\chi}{2} \p{\sp{1+\J_0\p{2\beta}} S_\z^2
  + \sp{1-\J_0\p{2\beta}} S_\y^2}.
\end{align}
Driving with a modulation index $\beta$ for which $J_0\p{2\beta}=\pm1/3$ then gives us the effective two-axis twisting Hamiltonians
\begin{align}
  H_{\t{eff}}^{(+)}
  &= \f{\chi}{3} \p{2 S_\z^2 + S_\y^2}
  \simeq \f{\chi}{3} \p{S_\z^2 - S_\x^2}, \\
  H_{\t{eff}}^{(-)}
  &= \f{\chi}{3} \p{S_\z^2 + 2 S_\y^2}
  \simeq \f{\chi}{3} \p{S_\y^2 - S_\x^2},
\end{align}
where $\simeq$ denotes equality up to the addition of a term proportional to $\v S^2=S_\z^2+S_\x^2+S_\y^2$, which is irrelevant in the absence of coherent coupling between states with different net spin.
In a similar spirit, one can work out that single-spin operators transverse to the $x$-axis transform as
\begin{multline}
  \tilde s_\pm
  \equiv \f12\p{-s_\z\pm i s_\y} \\
  \to U^\dag \tilde s_\pm U
  = e^{\pm i\beta\sin\p{\omega t}} \tilde s_\pm
  \approx \J_0\p{\beta} \tilde s_\pm,
\end{multline}
which implies that shifting into the rotating frame of the time-dependent drive takes
\begin{align}
  s_\x \to s_\x,
  &&
  s_{\y,\z} \to \J_0\p{\beta} s_{\y,\z},
  \label{eq:dec_transformation_xyz}
\end{align}
and
\begin{align}
  s_\pm
  \to \f12\sp{1 \pm \J_0\p{\beta}} s_+
  + \f12\sp{1 \mp \J_0\p{\beta}} s_-.
  \label{eq:dec_transformation_pm}
\end{align}
As the TAT Hamiltonians $H_{\t{eff}}^{(\pm)}$ are realized in a rotating frame, to properly account for decoherence throughout the TAT protocol one must transform jump operators according to Eqns.~\eqref{eq:dec_transformation_xyz}-\eqref{eq:dec_transformation_pm}.

In practice, our protocols realize the OAT Hamiltonian in Eqn.~\eqref{eq:driven_OAT} with additional $\sim\v S\c\v S$ and $\sim S_\z$ terms [see Eqn.~\eqref{eq:H_OAT}].
The effect of the $\sim\v S\c\v S$ term is to generate a relative phase between states with different total spin $S$ (with e.g.~$S=N/2$ within the Dicke manifold).
In the absence of coherent coupling between states with different total spin, therefore, the $\sim\v S\c\v S$ term has no effect on system dynamics.
The $\sim S_z$ term, meanwhile, is important; the magnitude of this term (as measured by the operator norm) is generally comparable to that of the squeezing term $\chi S_\z^2$.
Unlike in the case of OAT, $S_\z$ does not commute with the TAT Hamiltonians, so its effects cannot be eliminated by a single spin-echo $\pi$-pulse $\exp\p{-i\pi S_\x}$ half way through the squeezing protocol.
Nonetheless, we find that for $N=10^2$ ($10^3$) atoms, $\sim5$ (10) $\pi$-pulses in a CPMG (Carr-Purcell-Meiboom-Gill) sequence\cite{carr1954effects, meiboom1958modified} suffice to mitigate the effects of the $S_\z$ term in the TAT protocol (see Appendix \ref{sec:dynamical_decoupling}).
Phase control over these pulses, specifically choices of whether to apply $\exp\p{\pm i\pi S_\x}$ or $\exp\p{\pm i\pi S_\y}$ in any given $\pi$-pulse, can be used to construct XY-$n$ pulse sequences\cite{maudsley1986modified, gullion1990new} that are robust to pulse errors.

\section{Dynamical decoupling in the TAT protocol}
\label{sec:dynamical_decoupling}
\setcounter{figure}{0}

The effective Hamiltonian resulting from a perturbative treatment of SOC is (see Appendix \ref{sec:derivation_OAT})
\begin{align}
  H_{\t{eff}}
  = -\f{U}{L}\v S\c\v S - \overline B S_\z + \Omega S_\x + \chi S_\z^2,
  \label{eq:H_eff_DD}
\end{align}
where $U$ is a two-atom on-site interaction strength; $L$ is the number of lattice sites; $\overline B\equiv \sum_n B_n/N$ is a residual axial field determined by the occupied quai-momentum modes $\set{n}$ (with $\abs{\set{n}}=N$ atoms total); $\Omega$ is the magnitude of a driving field; and $\chi$ is an effective OAT squeezing strength.
The effect of the $\sim\v S\c\v S$ term is to generate a relative phase between states with different total spin $S$ (where $S=N/2$ within the Dicke manifold).
In the absence of coherent coupling between states with different total spin, therefore, the $\sim\v S\c\v S$ term has no effect on system dynamics, and we are safe to neglect it entirely.

In the parameter regimes relevant to our discussions in the main text, the operator norms of $\overline{B}\hat S_\z$ and $\chi \hat S_\z^2$ in Eqn.~\eqref{eq:H_eff_DD} will typically be comparable in magnitude.
The OAT protocol sets $\Omega=0$, and eliminates the effect of $\overline{B}\hat S_\z$ with a spin-echo $\pi$-pulse $\exp\p{-i\pi\hat S_\x}$ applied half way through the squeezing protocol.
The TAT protocol, meanwhile, effectively takes $\chi\hat S_\z^2+\Omega \hat S_\x\to\hat H_{\t{TAT}}^{(\pm)}$ (as defined in Appendix \ref{sec:derivation_TAT}) and $\overline{B}\hat S_\z\to\J_0\p{\beta_\pm}\overline{B}\hat S_\z$, where $\J_0$ is the zero-order Bessel function of the first kind and $\beta_\pm$ is the modulation index of the amplitude-modulated driving field $\Omega$, satisfying $\J_0\p{2\beta_\pm}=\pm1/3$.
Unlike in the case of OAT, $\hat S_\z$ does not commute with the TAT Hamiltonian, so its effect cannot be eliminated with a spin-echo.
Nonetheless, this term can be eliminated with a dynamical decoupling pulse sequence that periodically inverts the sign of $\hat S_\z$ while preserving $\hat H_{\t{TAT}}^{(\pm)}$.

\begin{figure*}
\centering
\includegraphics{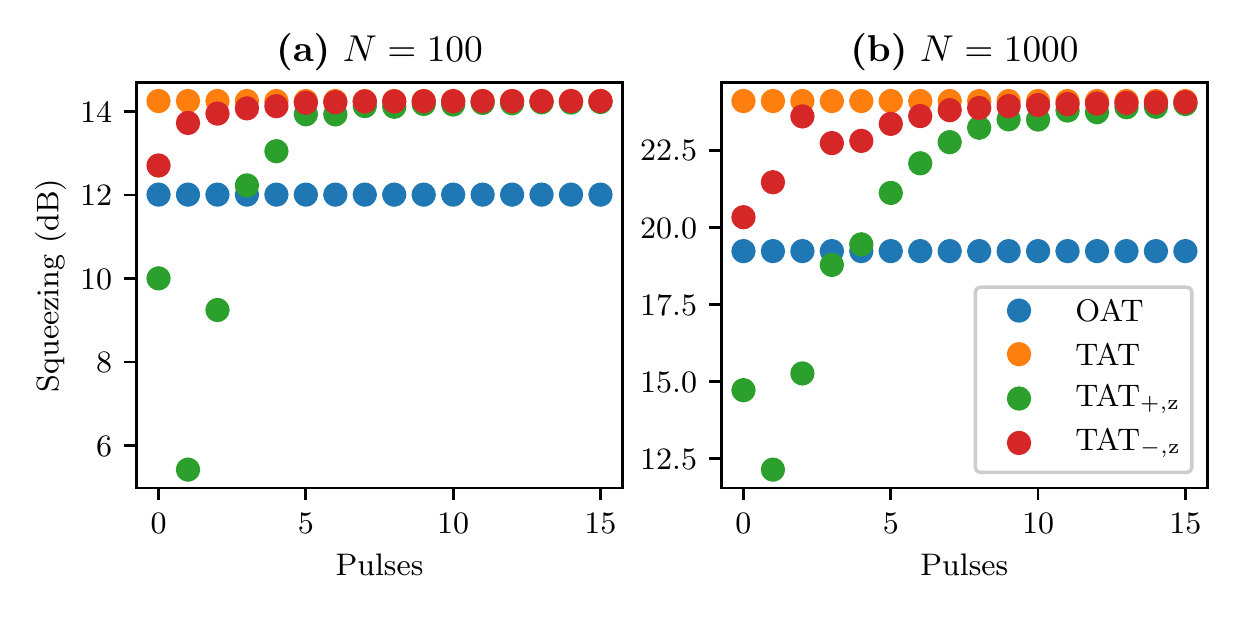}
\caption{{\bf Optimal squeezing as a function of $\pi$-pulses} applied prior to the optimal TAT squeezing time in a CPMG sequence with ({\bf a)} $N=100$ and ({\bf b}) $N=1000$ atoms.
Results are shown for OAT, TAT, and TAT$_{\pm,\z}$, where TAT$_{\pm,\z}$ denotes squeezing via the Hamiltonian $\hat H_{\t{TAT}}^{(\pm,\z)}\equiv \hat H_{\t{TAT}}^{(\pm)}-\J_0\p{\beta_\pm}\bk{\overline B}^{\t{rms}}_f \hat S_\z$.
Details about experimental parameters for these simulations are provided in the text.
}
\label{fig:pulsed_squeezing}
\end{figure*}

Fig.~\ref{fig:pulsed_squeezing} shows the maximal squeezing generated by $N=10^2$ and $10^3$ atoms via OAT, TAT, and TAT in the presence of the mean field $\J_0\p{\beta_\pm}\overline B\hat S_\z$ as a function of the number of $\pi$-pulses performed prior to the optimal TAT squeezing time.
These pulses are applied in a CPMG sequence $\p{\tau_n/2-\pi_\x-\tau_n/2}^n$, where $\tau_n/2$ denotes Hamiltonian evolution for a time $\tau_n/2$, $\pi_\x$ denotes the application of an instantaneous $\pi$-pulse $\exp\p{-i\pi\hat S_\x}$, and $n$ is the number of pulses, such that the optimal TAT squeezing time is $t_{\t{opt}}^{\t{TAT}}=\p{\tau_n}^n$.
The label TAT$_{\pm,\z}$ in Fig.~\ref{fig:pulsed_squeezing} denotes squeezing through the Hamiltonian $\hat H_{\t{TAT}}^{(\pm,\z)}\equiv \hat H_{\t{TAT}}^{(\pm)}-\J_0\p{\beta_\pm}\bk{\overline B}^{\t{rms}}_f\hat S_\z$, where $\bk{\overline B}^{\t{rms}}_f$ is the root-mean-square average of $\overline B$ over choices of occupied spacial modes $\set{n}$ at fixed filling $f$ of all spatial modes in the lowest Bloch band of a periodic 2D lattice.
While the modulation index $\beta_+$ is uniquely defined by $\J_0\p{2\beta_+}=1/3$, there are two choices of $\beta_-$ for which $\J_0\p{2\beta_-}=-1/3$; we use that which minimizes $\abs{\J_0\p{\beta_-}}$.
Fig.~\ref{fig:pulsed_squeezing} assumes an SOC angle $\phi=\pi/50$ (although results are independent of $\phi$ for $\phi\ll1$), a reduced field variance $\widetilde{B}/U=0.05$, and a filling $f=5/6$.
Note that as the filling $f\to1$, the residual axial field vanishes ($\overline{B}\to0$), so TAT$_{\pm,\z}\to$ TAT.

\section{Clock interrogation after squeezing}
\label{sec:clock_interrogation}

The protocols in our work concern the preparation of spin-squeezed states in an optical lattice clock.
Here, we discuss the use of these states in a follow-up clock interrogation protocol.
For simplicity, we restrict our discussion to the case of squeezing in 1D, as in Section \ref{sec:theory} of the main text, with the understanding that a generalization of this discussion to higher dimensions is straightforward.

A spin-squeezed state is generated by interactions and SOC that are generally undesirable during the clock interrogation protocol.
In the parameter regimes considered in our work, interactions alone have no effect on clock interrogation: absent of coherence between states with different net spin $S$, collective $\v S\c\v S$ interactions only generate unobservable global phases within each fixed-$S$ sector of Hilbert space.
Therefore, the remaining task to allow clock interrogation after spin squeezing is to turn off SOC, which inhomogeneously detunes atomic transition frequencies by an amount $B_q$ that depends on the quasi-momentum $q$ of an atom.
The SOC-induced axial fields $B_q\sim J\sin(\phi/2)$ depend on two tunable parameters: the tunneling rate $J$ and the SOC angle $\phi$.
The simplest way to turn off SOC is thus to increase the lattice depth prior to clock interrogation, taking $J\sim B_q\to0$.
Increasing the lattice depth to turn off SOC is compatible with the current clock interrogation sequence, but is incompatible with ongoing efforts to mitigate light scattering from the lattice beams (currently the primary source of decoherence in the clock; see Appendix \ref{sec:decoherence}) by using shallower lattices\cite{hutson2019engineering}.
We thus devote the rest of this section to discussing strategies for turning off SOC that are compatible with using the same lattice depth for clock interrogation as the spin squeezing generation.

If we cannot take the tunneling rate $J\to0$, our remaining control parameter for turning off SOC is the SOC angle $\phi=k_{\t{L}}a$, where $k_{\t{L}}$ is the projection of the clock laser wavenumber onto the lattice axis and $a$ is the lattice spacing.
The squeezing protocol needs a clock laser with a small but nonzero SOC angle $\phi\ll1$, while the clock interrogation protocol requires a clock laser with $\phi=0$.
Simply using one clock laser with $\phi\ne0$ for the squeezing protocol and another clock laser with $\phi=0$ for the clock interrogation protocol, however, does not resolve this discrepancy, because a state that is squeezed with respect to spin operators that are homogeneous (i.e.~of the form $S_\x,S_\y,S_\z$) in a particular gauge is not necessarily squeezed with respect to spin operators that are homogeneous in a different gauge.
In this appendix, we will work explicitly in the ``lab gauge'' of the clock interrogation protocol, in which the Fermi-Hubbard Hamiltonian is SOC-free and the $\phi=0$ clock laser is homogeneous.
To resolve the fact that our squeezing protocol prepares a state that is squeezed in the ``wrong gauge'', we will construct a simple pulse sequence that transforms the inhomogeneous spin operators accessible by the $\phi\ne0$ laser into a homogeneous form in our lab gauge.

Starting with a spin-down-polarized initial state
\begin{align}
  \ket{-\t{Z}}\equiv\p{\prod_j c_{j,\dn}^\dag}\ket{\t{vacuum}},
\end{align}
our OAT protocol prepares the state
\begin{align}
  \ket{\xi_{\t{OAT}}^{(\theta)}} = e^{-i H_{\t{FH}}^{(0)}t}
  e^{-i\p{\pi/2}S_\x^{(\theta)}} \ket{-\t{Z}},
\end{align}
where $H_{\t{FH}}^{(0)}$ is the Fermi-Hubbard Hamiltonian in Eqn.~\eqref{eq:FermiHubbard} without SOC; $t$ is some free evolution time; and $S_\x^{(\theta)}$ is the ``rotated'' spin-$x$-like generator induced by a clock laser with SOC angle $\phi=\theta$, namely
\begin{align}
  S_\x^{(\theta)}
  = \f12 \sum_j e^{i\theta j} c_{j,\up}^\dag c_{j,\dn} + \t{h.c.}.
\end{align}
Defining on-site spin operators (in the lab gauge)
\begin{align}
  s_\z^{(j)} &\equiv \f12 \p{ c_{j,\up}^\dag c_{j,\up}
    - c_{j,\dn}^\dag c_{j,\dn}} \\
  s_\x^{(j)} &\equiv \f12 \p{c_{j,\up}^\dag c_{j,\dn}
    + c_{j,\dn}^\dag c_{j,\up}} \\
  s_\y^{(j)} &\equiv \f{i}{2} \p{c_{j,\dn}^\dag c_{j,\up}
    - c_{j,\up}^\dag c_{j,\dn}},
\end{align}
we can identify the rotated collective spin operators
\begin{align}
  S_\x^{(\theta)} &\equiv \sum_j \p{\cos\p{\theta j} s_\x^{(j)}
  + \sin\p{\theta j} s_\y^{(j)}},
  \label{eq:S_x_theta} \\
  S_\y^{(\theta)} &\equiv \sum_j \p{\cos\p{\theta j} s_\y^{(j)}
  - \sin\p{\theta j} s_\x^{(j)}}.
  \label{eq:S_y_theta}
\end{align}
The state $\ket{\xi_{\t{OAT}}^{(\theta)}}$ is squeezed with respect to components of the rotated collective spin vector $\v S_\theta \equiv \p{S_\x^{(\theta)},S_\y^{(\theta)},S_\z}$.
Therefore, to take advantage of the squeezing in $\ket{\xi_{\t{OAT}}^{(\theta)}}$, the clock interrogation protocol effectively needs to rotate this state by some unitary $\exp\p{-i\v\eta\c\v S_\theta}$, and then extract information about the rotation vector $\v\eta$ from collective spin observables of the form
\begin{align}
  \bk{\O_\theta}_{\t{OAT}}^{\v\eta}
  \equiv \bk{\xi_{\t{OAT}}^{(\theta)}| e^{i\v\eta\c\v S_\theta}
    \O_\theta e^{-i\v\eta\c\v S_\theta} |\xi_{\t{OAT}}^{(\theta)}},
\end{align}
where $\O_\theta$ is some product of the rotated collective spin operators in $\v S_\theta$, e.g.~$S_\x^{(\theta)}$ or $S_\x^{(\theta)} S_\y^{(\theta)}$.
In order to turn off SOC during clock interrogation, however, we are restricted to performing rotations of the form $\exp\p{-i\v\eta\c\v S_0}$ and measuring homogeneous operators $\O_0$.
We thus seek a ``gauge-switching'' operation $G_\theta$ that maps homogeneous operators $\O_0$ onto rotated operators $\O_\theta$ via $G_\theta^\dag \O_0 G_\theta = \O_\theta$.
Equipped with $G_\theta$, we could decompose
\begin{align}
  \bk{\O_\theta}_{\t{OAT}}^{\v\eta}
  &= \bk{\xi_{\t{OAT}}^{(\theta)}| G_\theta^\dag e^{i\v\eta\c\v S_0}
    \O_0 e^{-i\v\eta\c\v S_0} G_\theta |\xi_{\t{OAT}}^{(\theta)}}
  \notag \\
  &= \bk{\tilde \xi_{\t{OAT}}^{(\theta)}| e^{i\v\eta\c\v S_0}
    \O_0 e^{-i\v\eta\c\v S_0} |\tilde \xi_{\t{OAT}}^{(\theta)}}
\end{align}
for a transformed state
\begin{align}
  \ket{\tilde \xi_{\t{OAT}}^{(\theta)}}
  \equiv G_\theta \ket{\xi_{\t{OAT}}^{(\theta)}}
\end{align}
that is now squeezed with respect to the homogeneous collective spin operators in $\v S_0$, accessible with the SOC-free ($\phi=0$) clock laser used during clock interrogation.

Given the definitions of the rotated spin operators $S_\x^{(\theta)},S_\x^{(\theta)}$ in Eqns.~\eqref{eq:S_x_theta}, \eqref{eq:S_y_theta}, a suitable candidate for the gauge-switching operator $G_\theta$ is the site-dependent rotation
\begin{align}
  G_\theta = \prod_j \exp\p{i\theta j s_\z^{(j)}}.
\end{align}
To implement $G_\theta$ with ``global'' (i.e.~non-site-selective) experimental controls, we decompose each local rotation into a product of two reflections:
\begin{align}
  \exp\p{i2\alpha s_\z^{(j)}}
  \simeq \exp\p{i\pi s_\x^{(j)}} \exp\p{i\pi s_\alpha^{(j)}}
\end{align}
where $\simeq$ denotes equality up to an overall phase, and $s_\alpha^{(j)} \equiv \cos\alpha s_\x^{(j)} + \sin\alpha s_\y^{(j)}$.
This decomposition implies
\begin{align}
  G_\theta \simeq \exp\p{i\pi S_\x^{(0)}} \exp\p{i\pi S_\x^{(\theta/2)}},
  \label{eq:G_pi_pi}
\end{align}
which can be implemented using one SOC-free ($\phi=0$) clock laser, and one clock laser with SOC angle $\phi=\theta/2$.
Appending the two $\pi$-pulses given by Eqn.~\eqref{eq:G_pi_pi} to our squeezing protocol thus prepares a state $\ket{\tilde\xi_{\t{OAT}}^{(\theta)}}$ that is squeezed with respect to the homogeneous collective spin operators $S_\x,S_\y,S_\z$ accessible to the SOC-free ($\phi=0$) clock laser used during clock interrogation.

As presented, the combined squeezing and clock interrogation protocols require three clock lasers in total: one without SOC ($\phi=0$), and one each for SOC angles $\phi\in\set{\theta/2,\theta}$.
We can use $G_\theta$, however, to decompose any pulse $\exp\p{-i\v\beta\c\v S_\theta}$ into composite pulses that use only $\v S_0$ and $\v S_{\theta/2}$:
\begin{align}
  \exp\p{-i\v\beta\c\v S_\theta}
  = G_\theta^\dag \exp\p{-i\v\beta\c\v S_0} G_\theta.
  \label{eq:pulse_decomposition}
\end{align}
The state prepared by the OAT squeezing protocol, for example, can be equivalently prepared via
\begin{align}
  \ket{\xi_{\t{OAT}}^{(\theta)}}
  &= e^{-i H_{\t{FH}}^{(0)}t} G_\theta^\dag
  e^{-i\p{\pi/2}S_\x^{(0)}} G_\theta \ket{-\t{Z}} \notag \\
  &\simeq e^{-i H_{\t{FH}}^{(0)}t} e^{-i\pi S_\x^{(\theta/2)}}
  e^{i\p{\pi/2} S_\x^{(0)}} \ket{-\t{Z}}.
\end{align}
Spin-echo pulses applied throughout OAT can likewise be decomposed according to \eqref{eq:pulse_decomposition}, eliminating the need for a clock laser with SOC angle $\phi=\theta$.
Applying a continuous drive during a squeezing protocol, however, still requires all three clock lasers.
If carefully tuning the relative orientations of three clock lasers proves to be too difficult in practice, converting OAT into TAT would therefore need to be done with a pulsed drive protocol, as in Ref.~[\citenum{liu2011spin}].

\section{Decoherence in the 3D $^{87}$Sr optical lattice clock}
\label{sec:decoherence}

Currently, light scattering from lattice beams in the 3D $^{87}$Sr optical lattice clock induces decoherence on a time scale of $\sim$10 seconds\cite{goban2018emergence, hutson2019engineering}.
This decoherence acts identically on all atoms in an uncorrelated manner, and can be understood by considering the density operator $\rho$ for a single atom, with effective spin states $\dn$ and $\up$ respectively corresponding to the ${}^1\t{S}_0$ and ${}^3\t{P}_0$ electronic states.
Empirically, the effect of decoherence after a time $t$ within the $\set{\dn,\up}$ subspace of a single atom is to take $\rho\to\rho\p{t}$ with $\rho\p{0}\equiv\rho$ and
\begin{align}
  \rho\p{t} \coloneqq
  \begin{pmatrix}
  \rho_{\up\up} e^{-\Gamma_{\up\up}t} &&
  \rho_{\up\dn} e^{-\Gamma_{\up\dn}t} \\
  \rho_{\up\dn}^* e^{-\Gamma_{\up\dn}t} &&
  \rho_{\dn\dn} + \p{1-e^{-\Gamma_{\up\up}t}} \rho_{\up\up}
  \end{pmatrix},
  \label{eq:decay_matrix}
\end{align}
where $\Gamma_{\up\up}\approx\Gamma_{\up\dn}\approx\Gamma=0.1~\t{sec}^{-1}$ are respectively decay rates for $\rho_{\up\up}$ and $\rho_{\up\dn}$.
This form of decoherence can be effectively modeled by decay and dephasing of individual spins (respectively denoted $\Gamma_{\t{ud}}$ and $\Gamma_{\t{el}}$ in Ref.~[\citenum{foss-feig2013nonequilibrium}]) at rates $\Gamma$.
In the language of the section that follows, we would say that this decoherence is captured by the sets of jump operators $\J_-\equiv\set{s_-^{(j)}}$ and $\J_\z\equiv\set{s_\z^{(j)}}$ with corresponding decoherence rates $\gamma_-=\gamma_\z=\Gamma$.

\section{Time-series of squeezing via OAT and TAT}
\label{sec:time_series}
\setcounter{figure}{0}

\begin{figure*}
\centering
\includegraphics{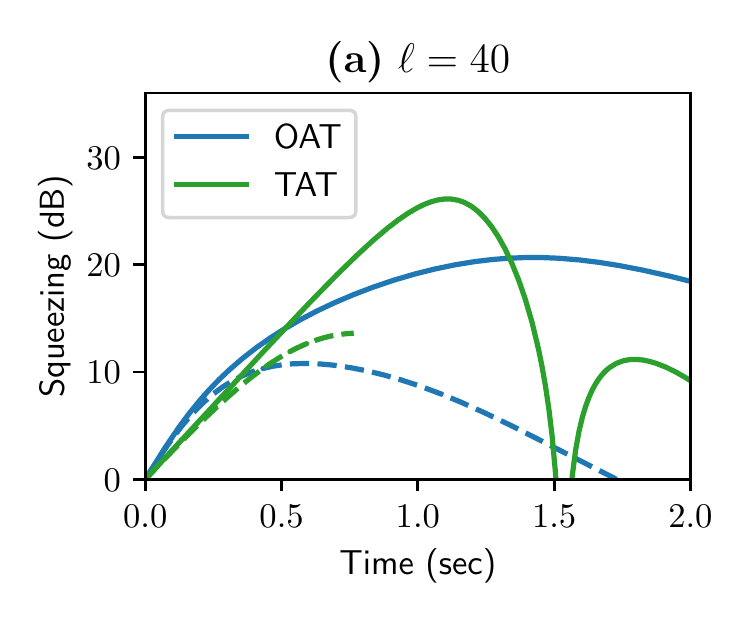}
\includegraphics{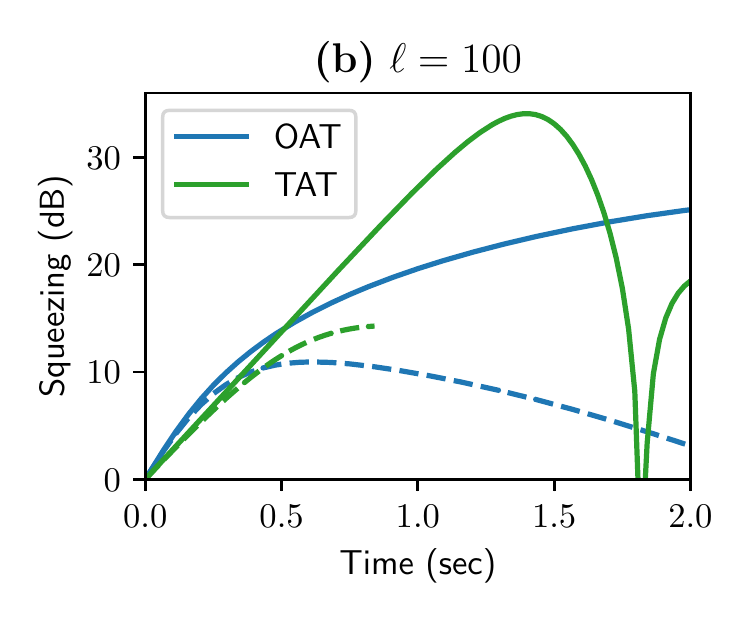}
\caption{{\bf Squeezing via OAT and TAT} in a 2D section of the 3D $^{87}$Sr optical lattice clock, shown for ({\bf a}) $\ell=40$ and ({\bf b}) $\ell=100$ sites per axis (with $N=\ell^2$ atoms total), and a lattice depth of $V_0=4~E_{\t{R}}$, where $E_{\t{R}}$ is the atomic lattice recoil energy.
Atoms are confined along the direction transverse to the 2D layer by a lattice of depth 60 $E_{\t{R}}$.
Squeezing over time is shown for OAT (blue) and TAT (green), both with (solid lines) and without (dashed lines) decoherence via uncorrelated decay and dephasing of individual spins at rates of $0.1~\t{sec}^{-1}$ (see Appendix \ref{sec:decoherence}).
}
\label{fig:squeezing_example}
\end{figure*}

Figure \ref{fig:squeezing_example} shows an example of squeezing over time via OAT and TAT, both with and without decoherence via decay and dephasing of individual spins.
The OAT model initially generates squeezing faster than the TAT model, but the squeezing generation rate of OAT (measured in dB per second) falls off with time.
The squeezing generation rate for TAT, meanwhile, remains approximately constant (in the absence of decoherence) until squeezing via TAT surpasses that of OAT.
In the absence of decoherence, OAT achieves a maximal amount of squeezing that scales as $\xi^2\sim N^{-2/3}$, while TAT achieves Heisenberg-limited squeezing with $\xi^2\sim N^{-1}$.
Note that our method for computing squeezing via TAT in the presence of decoherence (described in Appendix \ref{sec:collective_simulation}) is not capable of computing squeezing for the full range of times shown in Fig.~\ref{fig:squeezing_example}; the corresponding time-series data in this figure is therefore shown up to the point at which this method breaks down.

\section{Solving Heisenberg equations of motion for collective spin systems}
\label{sec:collective_simulation}

In order to compute squeezing of a collective spin system, we need to compute expectation values of (homogeneous) collective spin operators.
We compute these expectation values using a method recently developed in Ref.~[\citenum{perlin2019shorttime}], and provide a short overview of the method here.
Choosing the basis $\set{\S_{\v m}}$ for all collective spin operators, where $\S_{\v m}\equiv S_+^{m_+} S_\z^{m_\z} S_-^{m_-}$ with
$\v m\equiv\p{m_+,m_\z,m_-}\in\mathbb{N}_0^3$, we can expand all
collective spin Hamiltonians in the form
\begin{align}
  H = \sum_{\v m} h_{\v m} \S_{\v m}.
  \label{eq:H_general}
\end{align}
The evolution of a general correlator $\bk{\S_{\v n}}$ under a Hamiltonian of the form in Eqn.~\eqref{eq:H_general} is then given by
\begin{align}
  \f{d}{dt} \bk{\S_{\v n}}
  &= i \sum_{\v m} h_{\v m}
  \bk{\sp{\S_{\v m}, \S_{\v n}}_-}
  + \sum_\J \gamma_\J \bk{\mathcal D\p{\J; \S_{\v n}}} \notag \\
  &\equiv \sum_{\v m} \bk{\S_{\v m}} T_{\v m\v n},
\end{align}
where $\sp{X,Y}_\pm\equiv XY\pm YX$; $\J$ is a set of jump operators with corresponding decoherence rate $\gamma_\J$; the decoherence operator $\mathcal D$ is defined by
\begin{align}
  \mathcal{D}\p{\J;\O}
  \equiv \sum_{J\in\J}\p{J^\dag \O J - \f12\sp{J^\dag J,\O}_+};
\end{align}
and $T_{\v m\v n}$ is a matrix element of the time derivative operator $T\equiv d/dt$.
These matrix elements can be calculated analytically using product and commutation rules for collective spin operators.
We can then expand correlators in a Taylor series about $t=0$ to write
\begin{align}
  \bk{\S_{\v n}}
  &= \sum_{k\ge0} \f{t^k}{k!} \bk{\f{d^k}{dt^k} \S_{\v n}}_{t=0}
  \notag \\
  &= \sum_{k\ge0} \f{t^k}{k!}
  \sum_{\v m} T_{\v m\v n;k} \bk{\S_{\v m}}_{t=0},
  \label{eq:time_series}
\end{align}
where $T_{\v m\v n;k}\equiv\sp{T^k}_{\v m\v n}$ are matrix elements of the $k$-th time derivative.
Expectation values of collective spin operators can thus be computed via the expansion in Eqn.~\eqref{eq:time_series}, which at short times can be truncated at some finite order beyond which all terms have negligible contribution to $\bk{\S_{\v n}}$.

\section{Effect of a harmonic confining trap}
\label{sec:harmonic_trap}
\setcounter{figure}{0}

\begin{figure*}
\centering
    \includegraphics[width=0.7\linewidth]{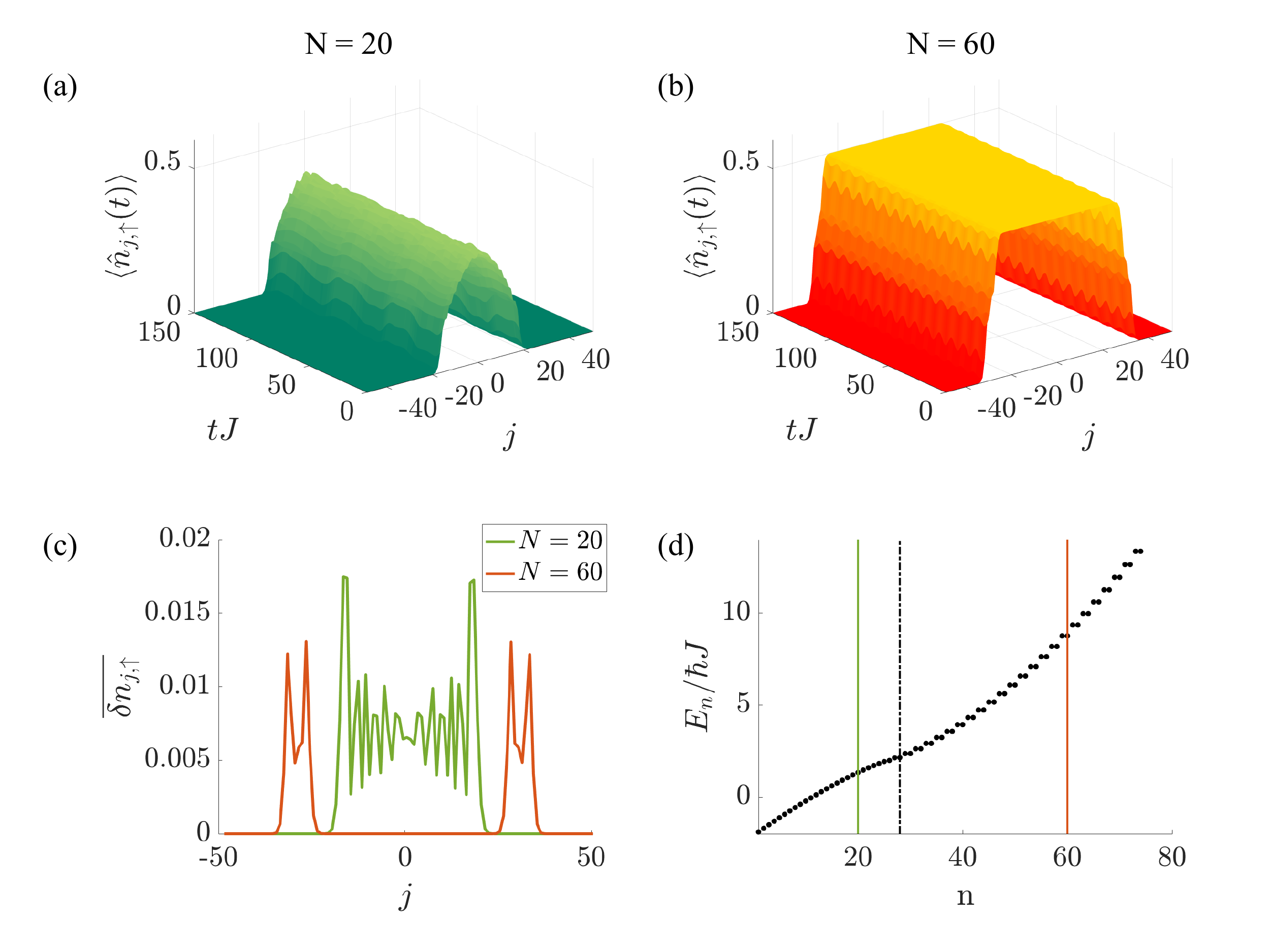}
    \label{fig:harmonic_trap1}
    \caption{Dynamics of non-interacting spin-orbit coupled fermions in a 1D lattice with SOC angle $\phi = \pi/50$, plus a harmonic trap with $\Omega/J = 0.01$. Starting with a spin-polarized cloud in $\downarrow$ ground state, an initial clock laser pulse is applied to rotate spins into $x$, and the atoms are allowed to evolve during the dark time. We track the dynamics of the $\uparrow$ particle density for the cases of (a) $N = 20$ and (b) $N = 60$ atoms. Panel (c) shows the time-averaged fluctuations of the $\uparrow$ particle density for each site index $j$ from its initial value following the Ramsey pulse; see Eqn.~\eqref{eq:time_avg_fluc}. For $N = 60$, we have filled all delocalized modes as well as several localized modes, resulting in a large region of no density fluctuations at the trap center. Panel (d) contains the eigenspectrum for a single internal state in the presence of the trap (with the index $n$ labeling the eigenvalues in order of increasing energy), where the critical mode $n_c$ dividing the spatially delocalized and localized modes is indicated  by a black dash-dotted line. The highest occupied mode in the $\downarrow$ ground state for $N = 20$ and $N = 60$ is indicated by the green and red solid lines, respectively.}
    \label{fig:harmonic_trap}
\end{figure*}

\begin{figure*}
    \includegraphics[width=0.95\linewidth]{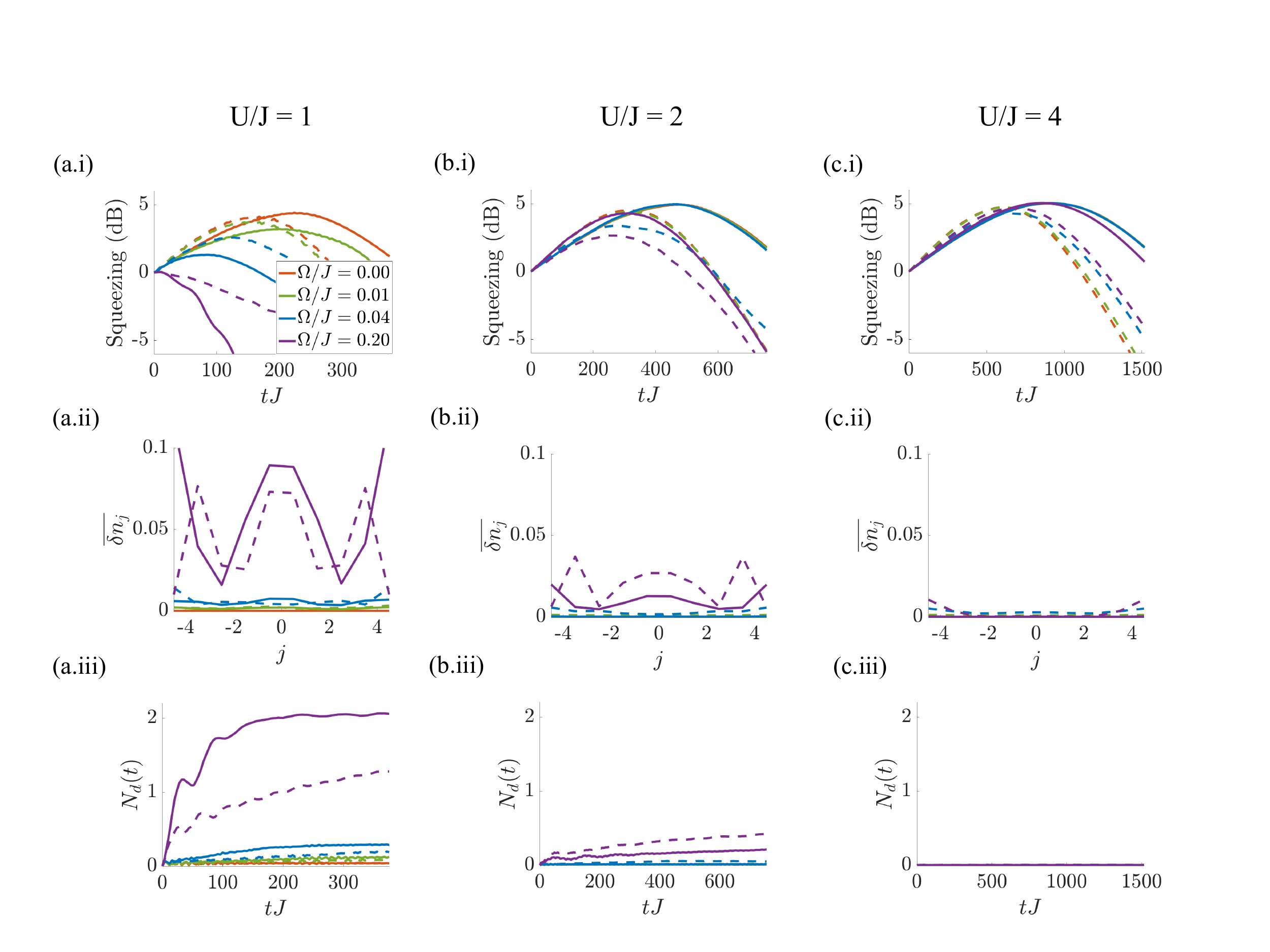}
    \label{fig:harmonic_trap_MB1}
    \caption{Dynamics of interacting spin-orbit  coupled fermions in a 1D lattice plus a harmonic trap for $U/J = 1$ (a), 2 (b), and 4 (c). For a 1D lattice with 10 sites and an SOC angle $\phi = \pi/50$, we apply a $\pi/2$ clock laser pulse to the $\downarrow$ ground state and let the system evolve during the dark time. In (a.i)-(c.i) we show the squeezing dynamics of the system for both $N = 10$ (solid lines) and $N = 9$ (dashed lines) for a variety of trapping strengths. In (a.ii)-(c.ii), we plot the time-averaged fluctuations in total particle density, $\overline{\delta n_j}$ (as in Eqn.~\eqref{eq:time_avg_fluc} but with $\hat{n}_{j,\uparrow}$ replaced by $\sum_{\alpha}\hat{n}_{j,\alpha}$). In (a.iii)-(c.iii), we plot the growth of the doublon population $N_d(t)$ (see Eqn.~\eqref{eq:doublon}) as a function of time, noting the absence of squeezing in the presence of a large doublon population. For the chosen trap strengths, the corresponding values of $n_c$ are 28 ($\Omega/J = 0.01$), 14 ($\Omega/J = 0.04$), and 6 ($\Omega/J = 0.2$). In panels where the results for the homogeneous case (orange curves) are not visible, they are nearly identical to the results for $\Omega/J = 0.01$ (green curves). Here, we utilize periodic boundary conditions to minimize finite size effects.}
    \label{fig:harmonic_trap_MB}
\end{figure*}

Current 3D optical lattice implementations involve a harmonic confining potential, which significantly alters the underlying single-particle eigenstructure and can potentially degrade accessible squeezing within our protocol. In this appendix, we examine the effect of a harmonic trap on our protocol and discuss strategies to mitigate undesired effects. We model the trap by the addition of the term
\begin{align}
    \hat{H}_{\Omega} = \hbar\Omega\sum_{j,\alpha}(j-j_0)^2\hat{n}_{j,\alpha}
\end{align}
to our Fermi-Hubbard model (Eqn.~\eqref{eq:FermiHubbard}), where $j_0$ denotes the trap center and $\Omega = m(\omega a)^2/2\hbar$ characterizes the trap strength for atom mass $m$, trap frequency $\omega$, and lattice spacing $a$. In current state-of-the-art 3D $^{87}$Sr OLC implementations, values of $\omega \approx 56 \times 2\pi~\t{sec}^{-1}$ can be achieved within each 2D layer of weak SOC by utilizing in-plane lattice depths of $5 E_R$ and a lattice depth of $60 E_R$ in the axial direction, resulting in a value of $\Omega/J \approx 0.01$. We restrict our discussion to 1D, although for a separable 3D lattice our arguments should extend in a straightforward manner.

We briefly review the structure of the single-particle eigenstates of the system, before discussing the effects on squeezing. In the quasi-momentum basis, the eigenfunctions $\psi_{n,\alpha}(q) = \braket{q|n,\alpha}$ are given by the $\pi$-periodic Mathieu functions, with the corresponding energies described by the Mathieu characteristic values\cite{rey2005ultracold}. In the presence of SOC, using  the gauge transformation described in the main text, we obtain the relation
\begin{gather}
    \psi_{n,\uparrow}(q) = \psi_{n,\downarrow}(q-\phi/a).
    \label{eq:trap_states}
\end{gather}
In contrast to the case of a pure harmonic potential, which generically has spatially delocalized single-particle eigenstates, the addition of a tight-binding lattice causes eigenmodes with quantum number $n$ (index $n$ labels the eigenvalues in order of increasing energy) larger than $n_c \approx 2\sqrt{2J/\Omega}$ to become localized at corresponding lattice sites. Therefore the  sites at a distance $n_c/2$ from the trap center with  potential energy $2\hbar J$ define  the boundary between the delocalized modes at the trap center and the high-energy localized trap edges. Tunneling in the region of  localized modes is typically suppressed by large potential energy differences even in the presence of SOC. These modes are thus largely decoupled and do not contribute to the trap center dynamics. On the other hand the delocalized modes may be approximated by those of a quantum harmonic oscillator with effective mass $m^{*} = \hbar/(2Ja^2)$ and frequency $\omega^* = \sqrt{4J\Omega}$.

As emphasized in the main text, the key requirements for our protocol are 1) the validity of the spin model, which depends on the pinning  of particles in their initial single particle modes, and 2) the gap protection against SOC dephasing, which arises from collective spin interactions. Concerning the latter point, it is desirable to maintain a weak trap so as to enable a large number of delocalized modes in the trap center, which are the only type capable of contributing to the generation of squeezing. Though the interactions between these modes are not strictly all-to-all, they remain long-ranged, and can thus still lead to a spin-locking effect and a protective gap\cite{rey2014probing, smale2019observation}. For $\Omega/J = 0.01$ we have $n_c = 28$, enabling $\sim10^3$ contributing modes in each 2D layer of our system. Concerning the validity of the spin model, from a single-particle perspective the eigenmodes  of our $\uparrow$ states will be initially displaced in quasi-momentum space from equilibrium by $\phi/a$ as per Eqn.~\eqref{eq:trap_states}, and will generally undergo dipole oscillations and not remain strictly pinned to their initial modes. However, as long as we ensure the displacement is small enough to guarantee  a constant density distribution across  the  trap center, the  spin model will remain valid. The localized modes at the trap edges can  actually help to satisfy this condition, since they can serve as a barrier against motion. This is demonstrated in Fig.~\ref{fig:harmonic_trap} where we show that filling all delocalized modes guarantees that the trap center maintains a constant density; we characterize this by the time-averaged fluctuations of the $\uparrow$ density at each site $j$ about its initial value following the Ramsey pulse,
\begin{align}
    \overline{\delta n_{j,\uparrow}} \equiv \sqrt{\lim_{t\to\infty}\frac{1}{t}\int_0^t \d\tau \bigg(\langle\hat{n}_{j,\uparrow}(\tau)\rangle - \langle\hat{n}_{j,\uparrow}(0)\rangle\bigg)^2},
    \label{eq:time_avg_fluc}
\end{align}
choosing sufficiently large evolution times to ensure convergence.

In the presence of interactions, an additional point of concern is that the interplay between the trap and interactions may induce resonances that enable the formation of a significant doublon population,
\begin{align}
    N_d(t) = \sum_j \langle\hat{n}_{j,\uparrow}(t)\hat{n}_{j,
    \downarrow}(t) \rangle,
    \label{eq:doublon}
\end{align}
which in turn may alter the density distribution and invalidate the spin model. Since doublon formation in the localized edges will not have consequences for our squeezing protocol, we must only ensure that doublons are not formed in the trap center, which requires $U > \Omega (n_c/2)^2 = 2J$\cite{pupillo2006extended}. In Fig.~\ref{fig:harmonic_trap_MB}, we perform exact simulations to assess the effect of the trap on our system. Though restricted to small system sizes, the results demonstrate that for $U/J \lesssim 2$, the trap will always lead to a decrease of squeezing due to the formation of doublons in the trap center, while for $U/J \gtrsim 2$, we are protected from this process even for trap strengths much stronger than the experimentally relevant ones.

\section{Accounting for $p$-wave inelastic collisions}
\label{sec:inelastic_collision}
\setcounter{figure}{0}

\begin{figure}[t]
\centering
\includegraphics[width=0.4\textwidth]{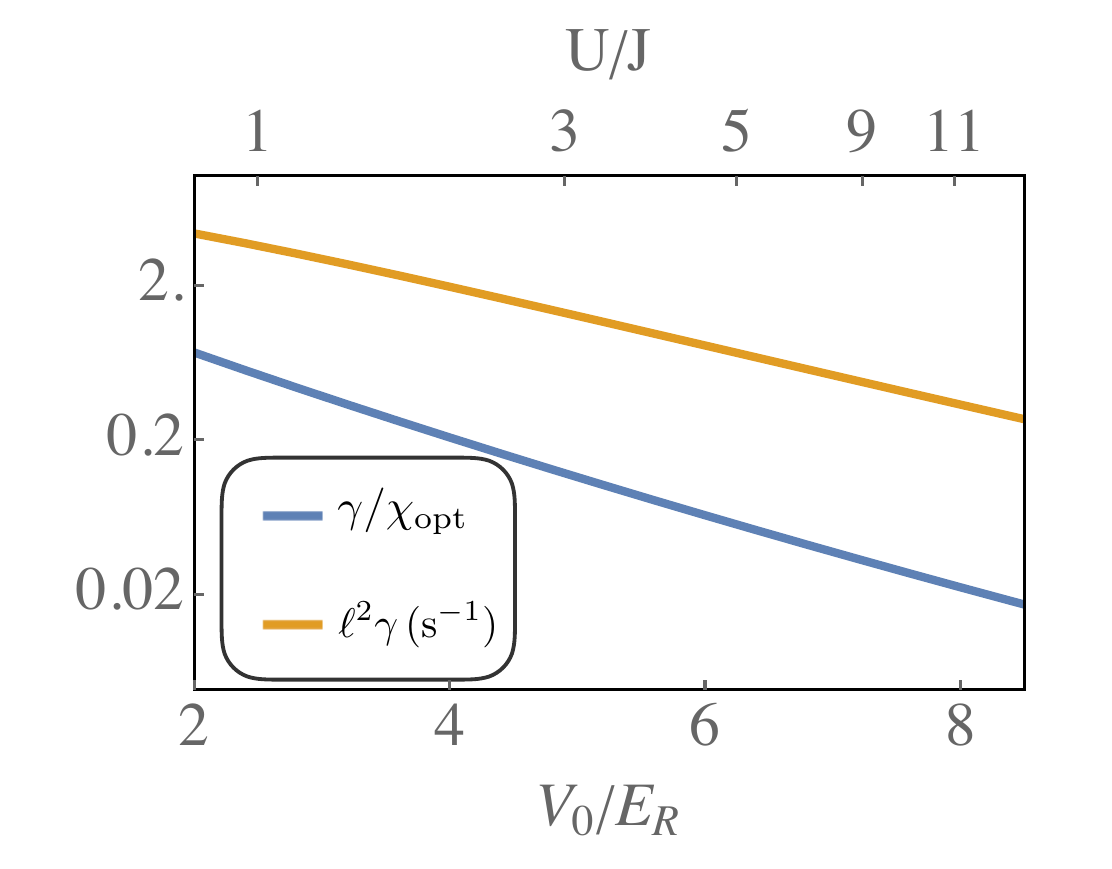}
\caption{
{\bf $p$-wave loss rates}.
Both the averaged $p$-wave inelastic collision rate $\gamma$ (orange) and the ratio of this collision rate to the optimal squeezing rate $\chi_{\text{opt}}$ (blue) are suppressed as the lattice depth increases.
$\chi_{\text{opt}}$ is obtained by choosing SOC angles $\phi$ that saturate $\widetilde{B}/U\approx0.05$, where $\widetilde{B}$ is the variance of the SOC-induced axial field and $U$ is the two-atom on-site interaction energy.
}
\label{fig:inelastic_rates}
\end{figure}

\begin{figure*}[t]
\centering
\includegraphics[width=0.7\textwidth]{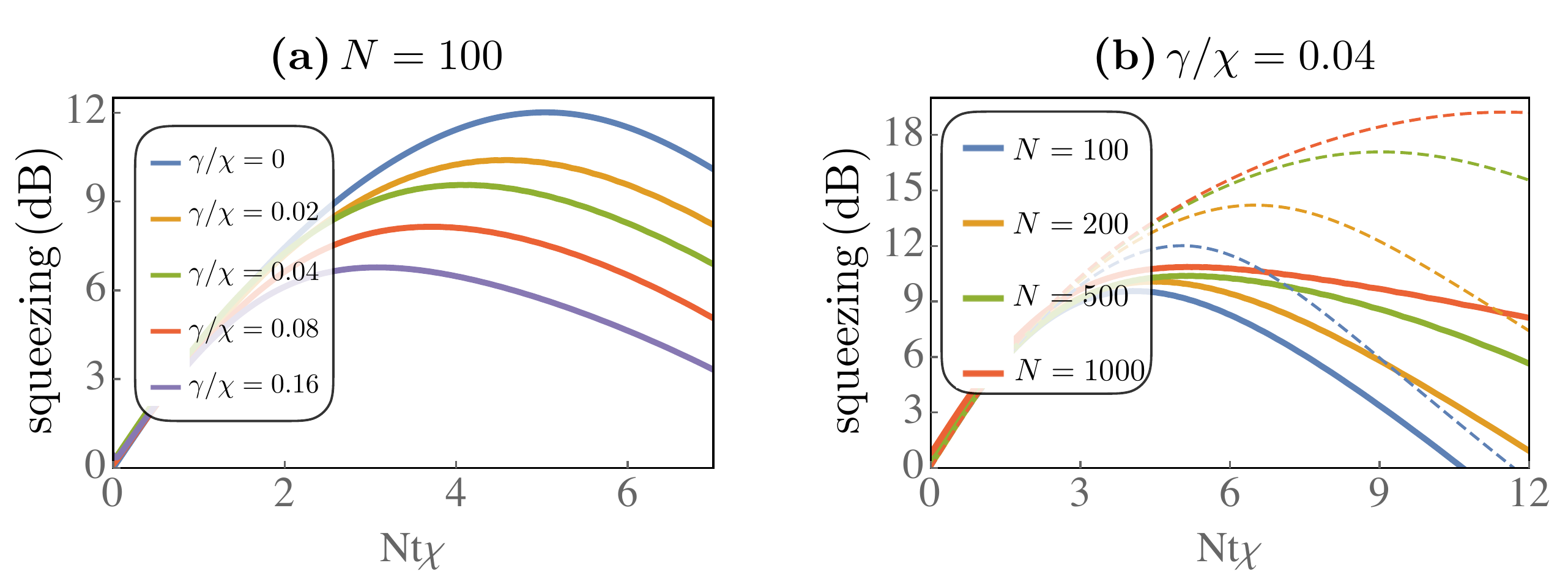}
\caption{\textbf{Squeezing via OAT in the presence of inelastic collisions}.
(a) For fixed particle number $N=100$, the optimal squeezing decreases as the inelastic collision rate increases.
Panel (b) shows squeezing over time for $\gamma/\chi_{\text{opt}}=0.04$ (solid lines), which corresponds to $U/J=6$,
and compares it with $\gamma=0$ (dashed lines) for  different particle numbers.
Inelastic collisions prevent the growth of optimal squeezing with particle number.
For $N=1000$, the maximum squeezing saturates to $\sim 10$ dB.
}
\label{fig:inelastic_squeezing}
\end{figure*}

Inelastic $^{3}\text{P}_{0}$ (electronic state $e$ or $\up$) collisions are detrimental for optical lattice clocks. For the  nuclear-spin-polarized gas discussed in this work, $ee$ losses are only possible via the $p$-wave scattering channel since $s$-wave collisions are suppressed by Fermi statistics. The big advantage here compared to  prior experiments done in a 1D lattice at $\mu$K temperature\cite{martin2013quantum} is that in a Fermi  degenerate  gas loaded in a 3D optical lattice, $p$-wave losses are further suppressed by the centrifugal barrier and Pauli blocking, and only happen through a wave-function overlap between atoms at different lattice sites.  In this appendix, we quantify the effect of $p$-wave interactions on squeezing. To account for $p$-wave losses, we describe the dynamics  using a master equation for the system's density matrix $\hat{\rho}$:
\begin{align}
  \frac{d\hat{\rho}}{dt}
  = -\frac{i}{\hbar} [\hat{H}_{\text{eff}},\,\hat{\rho}]
  + \mathcal{L}\hat{\rho},
\end{align}
where $\hat{H}_{\text{eff}}=\chi\hat{S}_z^2$ is the effective one-axis twisting Hamiltonian obtained from  the original Fermi-Hubbard Hamiltonian with spin-orbit coupling, and $\mathcal{L}$ is the Lindblad superoperator that accounts for $p$-wave $ee$ inelastic collisions. This latter term can be written using a pseudo-potential approximation as\cite{rey2014probing}:
\begin{multline}
  \mathcal{L}\hat\rho
  = \sum_{{\bm k}{\bm k'}} \Gamma_{{\bm k}{\bm k'}}
  \bigg[\hat{A}_{{\bm k}{\bm k'}}
  \hat\rho\hat{A}_{{\bm k}{\bm k'}}^\dag \\
  - \f12 \p{\hat{A}_{{\bm k}{\bm k'}}^\dag \hat{A}_{{\bm k}{\bm k'}} \hat\rho
+\hat\rho\hat{A}_{{\bm k}{\bm k'}}^\dag \hat{A}_{{\bm k}{\bm k'}}}\bigg],
\end{multline}
where the jump operators are $\hat{A}_{{\bm k}{\bm k'}}=\hat{c}_{{\bm k},\up}\hat{c}_{{\bm k'},\up}$, and
$\bm k$, $\bm k'$ sum over all the populated quasi-momentum modes. The decay matrix elements $\Gamma_{{\bm k}{\bm k'}}$ are given by:
\begin{align}
  \Gamma_{{\bm k}{\bm k'}}
  = \frac{3\pi \hbar b_{\text{im}}^3 }{m} \p{\int d{\bm r}^{\,3}
    W[\phi_{{\bm k}}({\bm r}),\phi_{{\bm k'}}({\bm r})]},
\end{align}
where $b_{\text{im}} = 121 a_0$\cite{zhang2014spectroscopic, goban2018emergence} is the $p$-wave inelastic scattering length (with $ a_0=5.29\times 10 ^{-11}$ m the Bohr radius), $\phi_{{\bm k}}({\bm r})$ is the Bloch function with quasi-momentum ${\bm k}$, and
\begin{multline}
  W\sp{\phi_{{\bm k}}({\bm r}),\phi_{{\bm k'}}({\bm r})} \\
  \equiv \sp{\p{{\bm \nabla}\phi^{*}_{{\bm k}}(\bm r)}
  \phi^{*}_{{\bm k'}}(\bm r)-\phi^{*}_{{\bm k}}(\bm r)
  \p{{\bm \nabla}\phi^{*}_{{\bm k'}}(\bm r)}} \\
  \cdot\sp{\p{{\bm \nabla}\phi_{{\bm k}}(\bm r)}
  \phi_{{\bm k'}}(\bm r)-\phi_{{\bm k}}(\bm r)
  \p{{\bm \nabla}\phi_{{\bm k'}}(\bm r)}}
\end{multline}

In Fig.~\ref{fig:inelastic_rates}, we show the averaged decay rate $\gamma\equiv\sum_{{\bm k}{\bm k'}}\Gamma_{{\bm k}{\bm k'}}/\ell^2$, where $\ell$ is the number of lattice sites along the $x$ and $y$ axes, as a function of the lattice depth $V_0$ along these axes.
Here, we assume the same lattice depth in the $z$ direction used in the main text, $V=60E_{\text{R}}$.
The decay rate $\gamma$ is suppressed exponentially with increasing lattice depth  $V_0$.
To quantify the effect of these losses on the spin squeezing generation process, we follow a similar methodology to the one described in detail in Ref.~\cite{rey2014probing}.
The basic idea is to take advantage of the so-called Truncated-Wigner Approximation (TWA)\cite{polkovnikov2010phase, schachenmayer2015manybody}, which allows us to capture the development of spin squeezing using semi-classical phase-space methods.
In the TWA the quantum dynamics are accounted for by solving mean field equations of motion supplemented by noise.
The mean field equations are  derived by assuming that the many-body density matrix of the system can be factorized as
$\hat{\rho}= \bigotimes_{i} \hat{\rho}({i})$, where $\hat{\rho}(i)$ is the reduced density matrix of the  particle in quasi-momentum  mode  ${\bm q}_i$ [see Eqn.~\eqref{eq:decay_matrix}].
Under this assumption, the non-linear mean field equations are given by
\begin{align}
  \frac{d\rho_{\up\up}(j)}{dt}
  = -\sum_{j'}\Gamma_{{\bm k}_j{\bm k}_{j'}}
  \rho_{\up\up}(j) \rho_{\up\up}(j'),
  &&
  \frac{d\rho_{\dn \dn}(j)}{dt}=0
\end{align}
and
\begin{multline}
  \frac{d\rho_{\up\dn}(j)}{dt}
  = \rho_{\up\dn}(j) \sum_{j'}
  \bigg[i\chi(\rho_{\up\up}(j')-\rho_{\dn\dn}(j')) \\
  - \f12 \Gamma_{{\bm k}_j{\bm k}_{j'}} \rho_{\up\up}(j')\bigg],
\end{multline}
where $\rho_{\sigma\sigma'}\equiv\langle\hat{\rho}_{\sigma\sigma'}\rangle$.
Since we are interested in the collective behavior, one can define $\rho_{\sigma\sigma'}^T=\sum_j \rho_{\sigma\sigma'}(j)$. For these observables the equations of motion simplify to
\begin{align}
  \frac{d\rho_{\up\up}^T}{dt} = -f\gamma  (\rho_{\up\up}^T)^2,
  &&
  \frac{d\rho_{\dn \dn}^T}{dt} = 0
\end{align}
and
\begin{align}
  \frac{d\rho_{\up\dn}^T}{dt}
  = \rho_{\up\dn}^T
  \sp{i\chi(\rho_{\up\up}^T-\rho_{\dn\dn}^T)
    - \f12 f\gamma \rho_{\up\up}^T},
\end{align}
where $f\equiv N/\ell^2$ is the filling fraction.

Under the TWA, one accounts for quantum fluctuations during the dynamics by  averaging over different mean field trajectories generated by sampling over different initial conditions chosen to reconstruct the Wigner function of the initial coherent spin state at $t=0$\cite{rey2014probing}.
This method has proven to be successful in simulating quantum spin dynamics.
Using this approach, Fig.~\ref{fig:inelastic_squeezing} shows numerical simulation results of squeezing over time in the presence of inelastic collisions.
For shallow lattices ($V_0\lesssim7E_{\t{R}}$), the effect of inelastic collisions can limit the spin squeezing to $\sim10$ dB.
Thus, in this regime, losses are as relevant as light scattering.
The role of inelastic interactions could be mitigated by either operating at deeper lattices as shown in Fig.~\ref{fig:inelastic_squeezing}, or by using nuclear spin states to generate the squeezing instead of the clock states directly.

\bibliography{main.bib}

\end{document}